\documentclass[journal,11pt]{IEEEtran}
\usepackage[final]{changes}
\usepackage{multirow}
\usepackage{makecell} % Auto-adjust columns for table
\usepackage[framed,autolinebreaks,useliterate]{mcode}%numbered
\usepackage{mathtools}
\usepackage{nicematrix}
\usepackage{float}
\usepackage{fancybox}
\usepackage{tabularx}
\usepackage{booktabs}
\usepackage[]{mdframed}
\usepackage{cancel}
\usepackage{enumitem} % to use (a) instead of 1)
\usepackage{ytableau}
\usepackage{graphicx} %package to manage images
\graphicspath{ {images/} }

\usepackage{color, colortbl}
\usepackage{multirow}

\usepackage{cuted}
\usepackage[linesnumbered,ruled,vlined]{algorithm2e} % ,noend %remove "vlined" for puting "end"

\usepackage[utf8]{inputenc}
\usepackage[english]{babel}
\usepackage{amsthm,amssymb}
\usepackage{amsmath}
\usepackage{bm}
\usepackage{optidef}
\usepackage{mathrsfs}

\usepackage[font=small,labelfont=bf,labelsep=space]{caption}
\usepackage{xcolor}
\def\BibTeX{{\rm B\kern-.05em{\sc i\kern-.025em b}\kern-.08em
    T\kern-.1667em\lower.7ex\hbox{E}\kern-.125emX}}

\usepackage[style=ieee,backend=biber]{biblatex}
\AtBeginBibliography{\footnotesize} %scriptsize

\usepackage{tikz}
\usetikzlibrary{shapes,matrix,backgrounds,positioning,arrows}

\usetikzlibrary{calc,patterns,decorations.pathreplacing,decorations.pathmorphing,matrix,positioning,arrows.meta,fit,bending}
\theoremstyle{definition}
\newtheorem{example}{Example}

\newtheorem{definition}{Definition}

\newcommand{\exampleqed}{\unskip\nobreak\hfill$\triangle$\par}
\let\oldexample\example
\let\oldendexample\endexample
\renewenvironment{example}
  {\oldexample}
  {\exampleqed\oldendexample}

\DeclareMathOperator{\bin}{bin}
\DeclareMathOperator{\supp}{supp}

\DeclareMathOperator{\w}{w}

\newcommand{\wm}{w_{\min}}

\newcommand{\Ki}{\mathcal{K}_i}
\newcommand{\A}{\mathcal{A}}

\newcommand{\I}{\mathcal{I}}

\renewcommand{\H}{\mathcal{H}}
\renewcommand{\S}{\mathcal{S}}

\newcommand{\J}{\mathcal{J}}

\newcommand{\M}{\mathcal{M}}

\newcommand{\MJ}{\mathcal{M}(\mathcal{J})}
\newcommand{\C}{\mathcal{C}}

\newcommand{\CI}{\mathcal{C}(\mathcal{I})}

\newcommand{\K}{\mathcal{K}}
\newcommand{\cO}{\mathcal{O}}

\newcommand{\bA}{\mathbf{A}}
\newcommand{\bB}{\boldsymbol{B}}

\newcommand{\bg}{\mathbf{g}}

\newcommand{\bb}{\mathbf{b}}

\newcommand{\bGN}{\boldsymbol{G}_N}
\newcommand{\bc}{\boldsymbol{c}}
\newcommand{\bu}{\boldsymbol{u}}

\newcommand{\bv}{\boldsymbol{v}}
\newcommand{\bw}{\boldsymbol{w}}

\newcommand{\bG}{\boldsymbol{G}}

\newcommand{\ind}{\operatorname{ind}}
\newcommand{\ev}{\operatorname{ev}}

\newcommand{\ft}{\mathbb{F}_2}

\definecolor{c1}{RGB}{250, 250, 250}
\definecolor{c2}{RGB}{250, 250, 200}
\definecolor{c3}{RGB}{250, 250, 150}
\definecolor{c4}{RGB}{250, 250, 100}
\definecolor{c5}{RGB}{250, 250, 50}
\definecolor{c6}{RGB}{250, 250, 1}
\definecolor{c7}{RGB}{250, 200, 1}
\definecolor{c8}{RGB}{250, 150, 1}
\definecolor{c9}{RGB}{250, 100, 1}
\definecolor{c10}{RGB}{250, 50, 1}
\definecolor{c11}{RGB}{250, 1, 1}
\definecolor{c12}{RGB}{200, 1, 1}
\definecolor{c13}{RGB}{150, 1, 1}
\definecolor{c14}{RGB}{100, 1, 1}
\definecolor{c15}{RGB}{50, 1, 1}

\definecolor{c16}{RGB}{212, 156, 142}
\definecolor{c17}{RGB}{180, 183, 224}
\definecolor{c18}{RGB}{212, 180, 224}
\newcommand{\Alow}{{\rm LTA}(m,2)}

\newcommand{\Ab}{(\mathbf{A},\mathbf{b})}

\newcommand{\Mon}{\mathcal{M}_{m}}

\DeclareFontFamily{U}{mathx}{}
\DeclareFontShape{U}{mathx}{m}{n}{<-> mathx10}{}
\DeclareSymbolFont{mathx}{U}{mathx}{m}{n}
\DeclareMathAccent{\widecheck}{0}{mathx}{"71}

\usepackage{tikz}
\definecolor{block}{RGB}{0,162,232}

\newcommand \yel[1]{\color{yellow}#1}
\newcommand \blue[1]{\color{block}#1}
\newenvironment{blockmatrix}{%
  \left(%
  \vcenter\bgroup\hbox\bgroup
  \tikzpicture[
    x=1.5\baselineskip,
    y=1.5\baselineskip,
  ]%
}{%
  \endtikzpicture
  \egroup
  \egroup
  \right)%
}

\newcommand*{\blocki}[1][block]{%
  \blockaux{#1}%
}
\def\blockaux#1(#2,#3)#4(#5,#6){%
  \draw[fill={#1}]
  let \p1=(#2,#3),
      \p2=(#5,#6),
      \p3=(#2+#5,#3+#6),
      \p4=(#2+#5/2,#3+#6/2)
  in
    (\p1) rectangle (\p3)
    (\p4) node {$#4$}
  ;%
}

\IEEEoverridecommandlockouts                              % This command is only
\title{
A Tutorial on Weight Structure of Polar Codes
}

\author{
\IEEEauthorblockN{Mohammad Rowshan and Vlad-Florin Drăgoi} %, and Jinhong Yuan$^\dagger$, {\em Fellow, IEEE}}
\thanks{\noindent Mohammad Rowshan %and Jinhong Yuan are 
is with the School of Electrical Engineering and Telecommunications, University of New South Wales (UNSW), Sydney, Australia, mrowshan@ieee.org.}
\thanks{\noindent Vlad-Florin Drăgoi is with Faculty of Exact Sciences, Aurel Vlaicu University, Arad, Romania, vlad.dragoi@uav.ro. His work was partially supported by RExQTCS: Romanian Excellence Quantum Technologies enhancing Cybersecurity, a grant of the Ministry of Education and Research from Romania, CCCDI - UEFISCDI, project number PN-IV-P6-6.1-CoEx-2024-0214, within PNCDI IV.}%, and with LITIS, University of Rouen Normandy, Saint-Étienne-du-Rouvray, France.}
\thanks{The authors contributed equally (Corresponding author: Vlad-Florin Drăgoi).}%, vlad.dragoi@uav.ro).}
}

\begin{document}

\maketitle
\thispagestyle{empty}
\pagestyle{empty}

%%%%%%%%%%%%%%%%%%%%%%%%%%%%%%%%%%%%%%%%%%%%%%%%%%%%%%%%%%%%%%%%%%%%%%%%%%%%%%%%

\begin{abstract}
%This tutorial provides an introduction to the algebraic foundations of polar codes' weight structure, exploring the formation and enumeration of codewords.
%{
This tutorial introduces the algebraic foundations underlying the weight structure of polar codes. Using a monomial-based polynomial formalism, we explain how polar and Reed–Muller codes can be viewed as decreasing monomial codes, enabling systematic characterization and enumeration of low-weight codewords. %Emphasis is placed on affine symmetries, orbit structures, and their role in weight distribution analysis. 
Its goal is to provide an accessible introduction to affine automorphisms, orbit-based descriptions of minimum and low-weight codewords, and their role in weight enumeration. Through illustrative examples and high-level overviews of recursive and coset-based techniques, the paper aims to prepare readers for deeper engagement with  the recent technical literature on the weight distribution of polar codes.
\end{abstract}

%\vlad{We might be required to add short bio!\\
%Also, we can drop the big Table with the counting formulae and have them included in the text where we explain and give examples. }
%%%%%%%%%%%%%%%%%%%%%%%%%%%%%%%%%%%%%%%%%%%%%%%%%%%%%%%%%%%%%%%%%%%%%%%%
\section{Introduction}

%Polar codes, introduced by Ar\i kan \cite{arikan}, are the first provably capacity-achieving codes with explicit construction that leverage channel polarization to transform identical channels into synthetic channels with varying reliability. %As short and medium-length polar codes have shown remarkable error correction performance, they had been chosen as a coding scheme for logical control channels in the fifth generation of the mobile broadband communication standard \cite{3GPP}. 
The algebraic properties of polar codes \cite{Arikan} can be effectively analyzed through polynomial formalism, which facilitates the characterization and enumeration of low-weight codewords. %This is paramount to better understanding of these codes and to paving the way to their applications, as it reveals the underlying mathematical structure governing their encoding, decoding, and error performance. 
By analyzing properties such as permutation (automorphism) groups, invariance under affine transformations, and decreasing order of generating monomials, researchers can improve decoding algorithms, identify subclasses of polar codes with enhanced performance, or assisting in weight enumeration using.% {closed-form formulae} or optimizing algorithmic methods. %Furthermore, a deeper algebraic understanding facilitates the integration of polar codes into new practical applications and opens avenues for hybrid coding schemes and postquantum cryptographic applications. 

The algebraic framework introduced in \cite{bardet} provided critical insights into the structure of polar codes, revealing key properties such as duality, permutation groups, and minimum-weight codewords. Shortened and punctured decreasing monomial codes were analyzed from a cryptographic perspective in \cite{bardet2016crypt,Dragoi-Szocs}. The permutation group plays an important role in polar coding research. An important subgroup of permutations, the lower triangular affine (LTA) group, was identified in \cite{bardet} {and later extended to the block lower triangular affine group in \cite{GEEB21}. 
In \cite{li2021complete}, it was shown that BLTA (and consequently LTA) constitute the only affine automorphisms of polar codes. Furthermore, \cite{IU22} demonstrated that in the asymptotic regime, the permutation group of polar codes converges to LTA group. %Among the most notable results on permutation group we list \cite{li2021complete, PBL22, IU22,IU22a, pillet2023distribution,IU22}. 
While BLTA were used in practical applications such as ensemble decoding \cite{GEECB21,PBL21,IU22}, LTA were used for structural properties, e.g., weight distribution. 

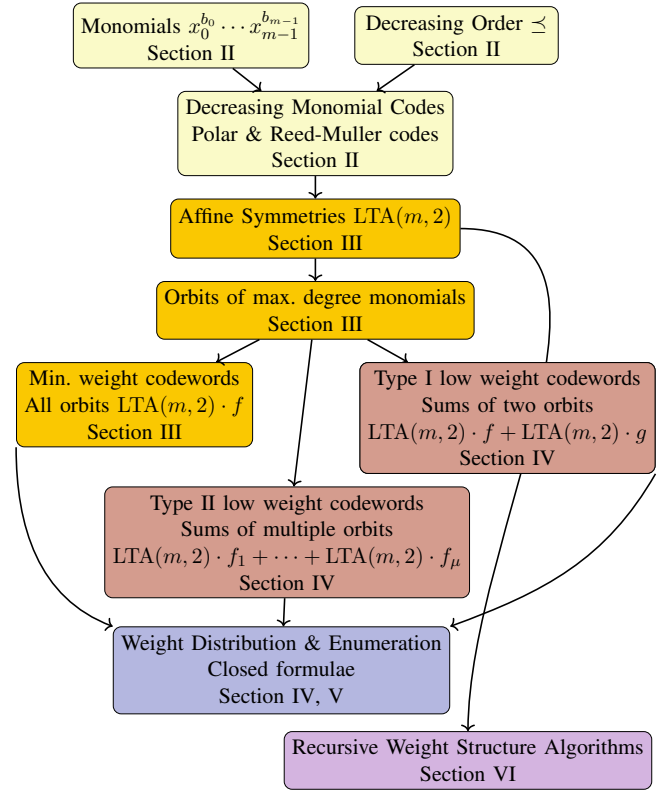
\begin{figure}[t]
\centering
\resizebox{0.48\textwidth}{!}{
\begin{tikzpicture}[
  node distance=0.4cm,
  every node/.style={draw, rectangle, rounded corners, align=center, minimum width=3.6cm, inner sep=4pt},
  arrow/.style={->, thick}
]

\node (m)[fill=c2] {Monomials $x_0^{b_0}\cdots x_{m-1}^{b_{m-1}}$\\Section \ref{sec:2}};
\node (d) [fill=c2,right=0.8cm of m] {Decreasing Order $\preceq$\\ Section \ref{sec:2}};
\node (dmc)[fill=c2,below=of m, xshift=2.2cm] {Decreasing Monomial Codes \\
Polar \& Reed-Muller codes\\Section \ref{sec:2}};
\node (a) [fill=c7,below=of dmc] {Affine Symmetries $\mathrm{LTA}(m,2)$\\  Section \ref{sec:3}};
\node (o) [fill=c7,below=of a] {Orbits of max. degree monomials\\ Section \ref{sec:3}};
\node (wmin) [fill=c7,below=0.6cm of o, xshift=-3.2cm,yshift=0.2cm] {Min. weight codewords \\ All orbits $\Alow\cdot f$\\Section \ref{sec:3}};
\node (wlowI) [fill=c16,below=0.6cm of o, xshift=3.4cm,yshift=0.2cm] {Type I low weight codewords\\ Sums of two orbits \\
$\Alow\cdot f+\Alow\cdot g$\\Section \ref{sec:wt_struct_enum}};
\node (wlowII) [fill=c16,below=1.2cm of o, xshift=-0.5cm, yshift=-1.4cm] {Type II low weight codewords \\ Sums of multiple orbits \\
$\Alow\cdot f_1+\dots+\Alow\cdot f_{\mu}$\\ Section \ref{sec:wt_struct_enum}};
\node (w) [fill=c17,below=0.5cm of wlowII, xshift=-0.1cm] {Weight Distribution \& Enumeration \\Closed formulae\\Section \ref{sec:wt_struct_enum}, \ref{sec:wt_short_code}};
\node (aw) [fill=c18,below=0.6cm of w,xshift=3.3cm,yshift=0.3cm] {Recursive Weight Structure Algorithms\\Section \ref{sec:recursive_enum}};

\draw[arrow] (m) -- (dmc);
\draw[arrow] (d) -- (dmc);
\draw[arrow] (dmc) -- (a);
\draw[arrow] (a) -- (o);
\draw[arrow] (o) -- (wmin);
\draw[arrow] (o) -- (wlowI);
\draw[arrow] (o) -- (wlowII);
\draw[arrow] (wmin.south west) to [out=270,in=160] (w.north west);
% Routed from south west to gracefully miss the newly moved 'aw' node
\draw[arrow] (wlowI.south east) to [out=240,in=20] (w.north east);
\draw[arrow] (wlowII) -- (w);

\draw[arrow] (a.east) to [out=0,in=90] (aw.north);

% Duplicate wlowI node to mask the intersecting arrow (drawn last, fill=white)
\node (wlowI_dup) [fill=c16, below=0.6cm of o, xshift=3.4cm,yshift=0.2cm] {Type I low weight codewords\\ Sums of two orbits \\
$\Alow\cdot f+\Alow\cdot g$\\Section \ref{sec:wt_struct_enum}};

\end{tikzpicture}
}
\caption{Conceptual roadmap of this tutorial.}
\label{fig:roadmap}\vspace{-10pt}
\end{figure}

%Recent studies have increasingly focused on the {\color{blue}entire} weight enumeration for polar codes. {Algorithmic approaches such as \cite{liu_analys,fossorier,polyanskaya,yao,liu2024method} are limited by complexity. 
%Recent studies have increasingly focused on the {\color{blue}entire} weight enumeration for polar codes. However, algorithmic approaches generally remain limited by high computational complexity. These works include early distance spectrum analyses \cite{liu_analys}, algorithms tailored for successive cancellation decoding \cite{fossorier,polyanskaya}, deterministic exact-weight computations \cite{yao}, and recent techniques aimed at reducing calculation overhead \cite{liu2024method}. 
While recent studies target the  weight enumeration for polar codes, algorithmic approaches remain limited by high computational complexity. These include distance spectrum analyses~\cite{liu_analys}, coset-wise weight spectrum evaluations~\cite{fossorier,polyanskaya}, exact-weight computations~\cite{yao}, and corresponding overhead-reduction techniques~\cite{liu2024method}. 
 Additionally, efficient algorithms evaluate the minimum weight ($w_{\mathrm{min}}$) of pre-transformed polar codes~\cite{rowshan2023minimum,ellouze2023low,ellouze2024computing,zunker2024enumeration} and partial weight distributions for punctured and shortened precoded codes~\cite{miroslavskaya2022}.
%{\color{blue} Also, efficient approaches were proposed for enumerating minimum-weight ($w_{\mathrm{min}}$) of pre-transformed polar codes in \cite{rowshan2023minimum,ellouze2023low,ellouze2024computing,zunker2024enumeration} as well as for computing the partial weight distribution of punctured, shortened precoded polar codes in \cite{miroslavskaya2022}.} %were proposed. New algorithms \cite{miroslavskaya2022} managed to enumerate all \vlad{CLARIFY:low-weight codewords} for polar codes with lengths up to 4096.} 
Closed-form formulae were also developed. In \cite{bardet}, a formula was derived for the number of $w_{\mathrm{min}}$ codewords in terms of orbits, i.e., actions of LTA on maximum degree monomials. %: $\Alow\cdot f.$   % based on the action of $\mathrm{A}_{\mathrm{low}}$. 
This was extended to $1.5w_{\mathrm{min}}$-weight codewords in \cite{vlad1.5d} and up to $2w_{\mathrm{min}}$-weight codewords in \cite{Ye2024-weightdistrib,rowshan2024weight}. {Most of the algorithmic and combinatorial approaches make use of the LTA.}

{
This tutorial does not introduce new theoretical results; its contribution lies in providing a clear, unified, and accessible exposition of recent advances. We begin with a monomial‑based framework that jointly describes polar and Reed–Muller codes as decreasing monomial codes. We then give an intuitive treatment of affine automorphisms, particularly the LTA group, and use their action to explain two types of low‑weight codewords, which cover all weights below twice the minimum distance. We also show how the monomial viewpoint clarifies the weight distribution of shortened polar codes and provide a high‑level overview of recursive and coset‑based methods for computing the weight spectrum. 
Fig. \ref{fig:roadmap} offers a conceptual roadmap. Numerous examples and accompanying code support experimentation and reproducibility, helping readers build familiarity with the key algebraic objects and techniques before engaging with more technical works such as \cite{bardet,rowshan2024weight,dragoi2025polar,dragoi2025weight,vlad1.5d}.
}
\section{From Binary Codes to Polynomials}\label{sec:2}

Binary codes are usually introduced as sets of binary vectors (codewords). However, there is another powerful way to think about codewords: as the output of a function evaluated at certain points. This viewpoint opens the door to representing codes using \emph{polynomial rings}. In this section, we show how a particular class of binary codes can also be described using polynomials, and how this idea helps us construct and understand powerful code families like polar codes and Reed-Muller codes. {The readers may find this section in more detail in \cite[Ch.~7]{RowshanViterbo2025}.}
%\subsection{Why Use Polynomial Rings in Coding Theory?}
%In binary coding theory, we typically work with vectors where each entry is 0 or 1. But there's another powerful way to think about codewords: as the output of a function evaluated at certain points. This viewpoint opens the door to representing codes using polynomial rings.%, which is especially useful for families like Reed-Muller and polar codes.

%%%%%%%%%%%%%%%%%%%%%%%%%%%%%%%%%%%%%%%%%%%%%%%%%%%%%%%%%%%%%%%%%%%%%%%%
\subsection{Polynomial Ring $\mathbb{F}_2\left[x_0, \ldots, x_{m-1}\right] /\left\langle x_i^2-x_i\right\rangle$}
We work over the field $\mathbb{F}_2$, which has only two elements: 0 and 1. In this field, addition is performed using the XOR operator. Now, $\mathbb{F}_2\left[x_0, \ldots, x_{m-1}\right]$ is the set of all polynomials you can form using the variables $x_0, \ldots, x_{m-1}$ with coefficients in $\mathbb{F}_2$. For example, you could have polynomials like $x_0$, $x_1 x_2$, or $x_0+x_1 x_3$, or even $x_0^3+x_1x_0^2$. A monomial is a single term, like $x_0 x_2 x_5$, where each variable appears at most once. The quotient $\left\langle x_i^2-x_i\right\rangle$ means we impose a rule on each variable: $x_i^2=x_i$ for every $i$. In $\mathbb{F}_2$, if $x_i=0$, then $x_i^2=0=x_i$, and if $x_i=1$, then $x_i^2=1=x_i$. This rule implies that %forces every variable to satisfy $x_i^2=x_i$, meaning 
higher powers like $x_i^3, x_i^4, \ldots$ are all equal to $x_i$. Effectively, each variable $x_i$ can only appear with exponent 0 or 1 in any monomial. Using the quotient, each variable acts like a binary switch (0 or 1), because $x_i^2=x_i$ implies $x_i\left(x_i-1\right)=0$, so $x_i=0$ or 1 when evaluated. This is useful for modeling binary systems. 
\subsection{Monomials Represent Boolean Functions}
Given a finite set of variables $\{x_0,x_1,\dots,x_{m-1}\}$, a monomial $f$ is simply a product of some variables $f=\prod_{j\in J}^{}x_{j}$, where $J$ is called the support of $f$ and denoted by $\ind(f).$ The monomial set used in the polar coding framework is defined as 
$$
\mathcal{M}_m \triangleq\left\{x_0^{b_0} \cdots x_{m-1}^{b_{m-1}} \mid\left(b_0, \ldots, b_{m-1}\right) \in \mathbb{F}_2^m\right\}.
$$
Let $f=x_0^{b_0} \ldots x_{m-1}^{b_{m-1}}\in\mathcal{M}_m$, the degree of $f$ is defined as $\deg(f)$, the Hamming weight of the exponent vector $\left(b_0, \ldots, b_{m-1}\right)$ or equivalently, the cardinality of $\ind(f)=\{x_i\mid b_i=1\}.$ In particular, $1$ is a monomial with degree zero. 
Each monomial corresponds to a basic Boolean function \cite{Carlet_2021}. 
%The representation used here is called the Algebraic Normal Form of a Boolean function. 
For example: 
\begin{itemize}
    \item 1 is the constant function (always 1),
    %\item $x_0$ returns the value of the variable $x_0$,
    \item $x_1 x_2$ is 1 only when both $x_1$ and $x_2$ are 1.
\end{itemize}

Any function $F(x_0, \ldots, x_{m-1})$ can be written as a sum of monomials, e.g., $
F(x_0, x_1, x_2)=x_0 x_1+x_2.$

%%%%%%%%%%%%%%%%%%%%%%%%%%%%%%%%%%%%%%%%%%%%%%%%%%%%%%%%%%%%%%%%%%%%%%%%
\subsection{Evaluation: From Functions to Codewords} 
To turn a polynomial (a function) into a codeword, we evaluate it at all possible binary inputs of length $m$. There are $2^m$ such inputs.
\begin{example}
    Take $m=3.$ %, so the code length equals $2^3=8$. 
% $$
% (0,0,0),(0,0,1),(0,1,0),(0,1,1),(1,0,0),(1,0,1),(1,1,0),(1,1,1)
% $$
Let's evaluate $F\left(x_0, x_1, x_2\right)=x_0 x_1+x_2$ at all input combinations of $(x_0,x_1,x_2)$ for $x_i\in\{0,1\}$ (also called points).%\vlad{Use decreasing order}
\begin{center}
\footnotesize
\begin{tabular}{|c|c|c||c|c|}
\hline \rowcolor{c2}$x_0$ & $x_1$ & $x_2$ & $x_0x_1$ & $F\left(x_0, x_1, x_2\right)$ \\
\hline 0 & 0 & 0 & 0 & 0 \\
\hline 1 & 0 & 0 & 0 & 0 \\
\hline 0 & 1 & 0 & 0 & 0 \\
\hline 1 & 1 & 0 & 1 & 1 \\
\hline 0 & 0 & 1 & 0 & 1 \\
\hline 1 & 0 & 1 & 0 & 1 \\
\hline 0 & 1 & 1 & 0 & 1 \\
\hline 1 & 1 & 1 & 1 & 0 \\
\hline
\end{tabular}
\end{center}
{ We order the set of elements in $\ft^m$ by sorting in decreasing order the corresponding integers. Considering the binary expansion of all integers from $2^m-1$ down to $0$, i.e., $\bin(i)=(i_0,\dots, i_{m-1})$ satisfying $i=\sum_{j=0}^{m-1}i_j2^j$, we obtain an ordered set of evaluation points. Gathering all values of $F(x_0,x_1,x_2)$ in order we have
%So, the resulting codeword corresponds to the evaluation of $f=x_0 x_1+x_2$, denoted by $\ev(f)$:
$$\footnotesize
\begin{array}{lcccccccc}
i &7 & 6 & 5&4&3&2&1&0\\
\bin(i) & 111 & 011 & 101 & 001 & 110 & 010 & 100 & 000 \\
\ev(F)=\quad &\text{[ }0 & 1 & 1 & 1 & 1 & 0 & 0 & 0\text{ ]}
\end{array}
$$
}
%where the elements are placed in reverse order of the table above, aligning with the polar codes literature. %The codeword $01101010$ is thus the evaluation $\ev(f).$
\end{example}
%\vlad{Maybe first talk about $\bG_2$}

{
Let us illustrate how evaluation of monomials maps into polar codes construction, using the binary kernel of the polar codes. The simplest case is for $m=1$ where the kernel $\bG_{2}=\begin{pmatrix}
    1&0\\
    1&1
\end{pmatrix}$ is the evaluation of monomial $x_0$ (first row) and monomial $1$ (second row) over the points $(1),(0).$ This can be extended to 3 variables for example, by taking the Kronecker product of $\bG_{2}$ with itself 3 times, as follows
{\footnotesize
\[\bG_{2^2}=\bG_2\otimes\bG_2=\begin{blockmatrix}
    \blocki(0,0)\bG_2(1.2,1.2)
    \blocki(1.2,0)\bG_2(1.2,1.2)
    \blocki(0,1.2)\bG_2(1.2,1.2)
    \blocki[yellow](1.2,1.2)0(1.2,1.2)
  \end{blockmatrix}=
  \begin{pmatrix}
  \begin{array}{cc|}\blue{1} & \blue{0} \\ 
                                   \blue{1} & \blue{1} 
\end{array}
&\begin{array}{cc} \yel{0} & \yel{0} \\  
                                     \yel{0} & \yel{0} 
\end{array} \\\cline{1-2} 
  \begin{array}{cc|} \blue{1} & \blue{0}\\ 
  				\blue{1} & \blue{1}
\end{array}
& \begin{array}{cc}\blue{1} & \blue{0} \\ 
			 \blue{1} &\blue{1}
\end{array}
\end{pmatrix} \]

\[\bG_{2^3}=\bG_2\otimes\bG_{2^2}=\begin{blockmatrix}
    \blocki(0,0)\bG_{2^2}(1.2,1.2)
    \blocki(1.2,0)\bG_{2^2}(1.2,1.2)
    \blocki(0,1.2)\bG_{2^2}(1.2,1.2)
    \blocki[yellow](1.2,1.2)0(1.2,1.2)
  \end{blockmatrix}=\begin{blockmatrix} 
   \blocki(0,0)\bG_2(1,1)
    \blocki(1,0)\bG_2(1,1)
    \blocki(0,1)\bG_2(1,1)
    \blocki[yellow](1,1)0(1,1)
   
   \blocki(2,0)\bG_2(1,1)
    \blocki(3,0)\bG_2(1,1)
    \blocki(2,1)\bG_2(1,1)
    \blocki[yellow](3,1)0(1,1)
   
     \blocki(0,2)\bG_2(1,1)
    \blocki(1,2)\bG_2(1,1)
    \blocki(0,3)\bG_2(1,1)
    \blocki[yellow](1,3)0(1,1)

   \blocki[yellow](2,2)0(1,1)
    \blocki[yellow](3,2)0(1,1)
    \blocki[yellow](2,3)0(1,1)
    \blocki[yellow](3,3)0(1,1)
   
  \end{blockmatrix}\]
  \[\bG_{2^3}=\begin{pmatrix}
  \begin{array}{cc|}\blue{1} & \blue{0} \\ 
                                   \blue{1} & \blue{1} 
\end{array}
&\begin{array}{cc|} \yel{0} & \yel{0} \\  
                                     \yel{0} & \yel{0} 
\end{array} 
&\begin{array}{cc|} \yel{0} & \yel{0} \\  
                                     \yel{0} & \yel{0} 
\end{array}
&\begin{array}{cc} \yel{0} & \yel{0} \\  
                                     \yel{0} & \yel{0} 
\end{array}
\\\cline{1-4} 
  \begin{array}{cc|} \blue{1} & \blue{0}\\ 
  				\blue{1} & \blue{1}
\end{array}
& \begin{array}{cc|}\blue{1} & \blue{0} \\ 
			 \blue{1} &\blue{1}
\end{array}
&\begin{array}{cc|} \yel{0} & \yel{0}\\  
                                     \yel{0} & \yel{0} 
\end{array}
&\begin{array}{cc} \yel{0} & \yel{0} \\  
                                     \yel{0} & \yel{0} 
\end{array}
\\\cline{1-4}
\begin{array}{cc|}\blue{1} & \blue{0} \\ 
                                   \blue{1} & \blue{1} 
\end{array}
&\begin{array}{cc|}\yel{0} & \yel{0} \\  
                                     \yel{0} & \yel{0} 
\end{array} 
&\begin{array}{cc|}\blue{1} & \blue{0} \\ 
                                   \blue{1} & \blue{1} 
\end{array}
&\begin{array}{cc} \yel{0} & \yel{0} \\  
                                     \yel{0} & \yel{0} 
\end{array} 
\\\cline{1-4}
  \begin{array}{cc|} \blue{1} & \blue{0}\\ 
  				\blue{1} & \blue{1}
\end{array}
& \begin{array}{cc|}\blue{1} & \blue{0} \\ 
			 \blue{1} &\blue{1}
\end{array}
&  \begin{array}{cc|} \blue{1} & \blue{0}\\ 
  				\blue{1} & \blue{1}
\end{array}
& \begin{array}{cc}\blue{1} & \blue{0} \\ 
			 \blue{1} &\blue{1}
\end{array}
\end{pmatrix} 
\begin{array}{r}
x_0x_1x_2\\
x_1x_2\\
x_0x_2\\
x_2\\
x_0x_1\\
x_1\\
x_0\\
1
\end{array}
\]
}
}Likewise, we can find the rows of the matrix $\bG_{2^3}$ by evaluating all monomials up to degree $m=3$ using the following mapping between row indices and monomials:

$i \in [0,2^m\!-\!1] \leftrightarrow \bin(i) \in \ft^m \leftrightarrow \supp\left(\bin(i)^c\right) =\ind(f) \leftrightarrow f\in\Mon$, where $\bin(i)^c=\bin(2^m\!-\!1\!-\!i).$ Each row index  corresponds to a monomial, e.g., $\bin(1)=(1,0,0)\rightarrow \bin(1)^c=(0,1,1)\rightarrow f=x_1x_2.$ Hence, monomials represent rows of $\bG_{2^m}$ and polynomials correspond to rows summations.% of rows of this matrix.

{ We can define divisibility between monomials, i.e., for any pair of monomials $f$ and $g$ we write $f|g$ iff $\ind(f)\subseteq\ind(g).$ For example, $x_0|x_0x_1.$ Using divisibility we can now define the concept of \emph{multiplicative complement} of a monomial. We write $\widecheck{f}=(x_0\dots x_{m-1})/f$ to denote this concept,} e.g., for $m=3$, $\widecheck{x_0x_1x_2}=1$, and $\widecheck{x_1}=x_0x_2.$ %Complement can also be defined using the set of variables of a monomial $\ind(f)$, and $\ind({\widecheck{f}})\!=\!\{0,\dots,m\!-\!1\}\!\setminus\! \ind({f}).$

%%%%%%%%%%%%%%%%%%%%%%%%%%%%%%%%%%%%%%%%%%%%%%%%%%%%%%%%%%%%%%%%%%%%%%%%
\subsection{Hamming Weight of Polynomials}
%The Hamming weight of a binary vector is the number of positions where the vector has a 1. 
%Since a polynomial $F\left(x_0, \ldots, x_{m-1}\right)$ defines a Boolean function, evaluating $F$ at all $2^m$ binary inputs gives a vector (codeword) of length $2^m$. 
{Let $F\left(x_0, \ldots, x_{m-1}\right)$ be a polynomial and $\ev(F)$ the corresponding codeword.} The Hamming weight of $\ev(F)$ is the number of inputs $u \in \mathbb{F}_2^m$ such that $F(u)=1$
$$
\w(\ev(F))= |\{u \in \mathbb{F}_2^m \mid F(u)=1\}|.
$$

Suppose $F$ can be written as a product of independent linear forms, i.e., $F=l_1\cdots l_r$ where $\deg(l_i)=1$ and we can not express any $l_i$ as a linear combination of the other linear terms $l_j.$ For example $F=(x_1+x_0)(x_2+1)$ is a product of two independent linear forms since both $x_1+x_0$ and $x_2+1$ are linear (of degree 1), and independent. For such $F$ we have %satisfies such a condition then % the monomial is $x_{i_1} x_{i_2} \cdots x_{i_r}$ (with $r$ distinct variables), then:
$
\w(\ev(F))=2^{m-r}
$ (\cite{macwilliams,Carlet_2021})
in particular, 
%\begin{equation}
 $   \w(\ev(x_{i_1}x_{i_2}\dots x_{i_{r}}))=2^{m-r}.$
%\end{equation}
% \vlad{These have to be products of independent linear forms and we need to mention what independent linear forms are}
% One can find a simple proof of this result in \cite{Carlet_2021}. 
% \begin{example}
% For $f=x_0 x_1 x_2$, the degree is 3 and $m=4$. Then:
% $$
% w(f)=2^{4-3}=2^1=2
% $$
% Only two inputs out of 16 will satisfy $x_0=x_1=x_2=1$, where $x_3\in\{0,1\}$.
% \end{example}
% \vlad{This is well-known result from \cite{kasami1970weight} on product of independent linear forms, and consequently on product of distinct variables. Maybe we should add some explanations on linear forms, independence.}
%%%%%%%%%%%%%%%%%%%%%%%%%%%%%%%%%%%%%%%%%%%%%%%%%%%%%%%%%%%%%%%%%%%%%%%%
\subsection{From Monomials to Generator Matrices}
%  The $\ev$ operator allows us to define polynomial/monomial codes. Indeed, any $\I \subseteq \M_{m}$ forms a \emph{generating set} for a ($n=2^m, k=|\I|$) monomial code $\C$.  As a linear code $\CI=\operatorname{span}(\{\ev(f) \mid f \in \I\}),$
% where $\ev(f)=\left(\ev_{\bz}(f) : \bz\in \ft^m\right)$ is the binary vector obtained by evaluating  $f$ over all the binary entries in $\ft^m$ (see \cite{dragoi17thesis}). Here, the elements of $\ft^m$ are ordered using the index order, where $z_{m-1}$ is the most significant bit of $\bz=(z_0,\dots,z_{m-1})\in\ft^m.$

{Given a monomial set $\mathcal{I} \subseteq \M_m$, a code can be constructed by evaluating each $f\in\I$ using the $\ev()$ function. Stacking these rows gives the generator matrix $\bG$ of the linear code $\C$ of length $2^m$ and dimension $|\I|.$  
Every monomial in $\I$ can be mapped to row index $i$ of the matrix $\bG_{2^m}$ that is collected in the set $\A$. %This is equivalent to computing the row-submatrix of $\bGN$ corresponding to $\I.$ 
}

% Given a monomial set $\mathcal{I} = \{g_1, \ldots, g_k\} \subseteq \M_m$, a code can be constructed by evaluating each monomial at all $2^m$ binary input vectors. Each monomial $g_i \in \mathcal{I}$ gives a row of the generator matrix $G$ via:
% $$
% F_{i,:}=\left[g_i\left(u_{2^m-1}\right)\; g_i\left(u_{2^m-2}\right)\; \ldots\; g_i\left(u_0\right)\right],
% $$
% where $u_j \in \mathbb{F}_2^m$ are the binary inputs ordered lexicographically. Stacking these rows gives the generator matrix $G$. 
% Every monomial $f\in\I$ can be mapped to row index $i$ of $F_{2^m}$ that is collected in the set $\A$. 

\begin{example}
Polar code of length 8 (i.e., $m=3$).
Let us choose the following 4 monomials as the generating set:
$$
\mathcal{I}=\left\{{x_2}, x_1, x_0,1\right\},\;\;\A=\left\{3,5,6,7\right\}.
$$
By evaluating the monomials over all 8 binary input vectors, we obtain the rows of the generator matrix $\bG$ of size $4 \times 8$. The corresponding code is defined as the set of all linear combinations of these rows.
% Hence, 
% $$
% \bG\!=\!\left[\begin{array}{cccccccc}
% \ev\left(x_0 x_1\right) \\
% \ev\left(x_1\right) \\
% \ev\left(x_0\right) \\
% \ev(1)
% \end{array}\right]\!=\!
% \left[\begin{array}{cccccccc}
% 1 & 0 & 0 & 0 & 1 & 0 & 0 & 0  \\
% 1 & 1 & 0 & 0 & 1 & 1 & 0 & 0  \\
% 1 & 0 & 1 & 0 & 1 & 0 & 1 & 0  \\
% 1 & 1 & 1 & 1 & 1 & 1 & 1 & 1 
% \end{array}\right]
% %\Rightarrow \text { generator matrix for a length-8, dimension-4 polar code. }
% $$
\end{example}

This is how Reed-Muller and polar codes are built: (a) Reed-Muller code of order $r$ : take all monomials of degree $\leq r$. (b) Polar code: choose a subset of monomials (not necessarily by degree) that correspond to reliable bit-channels in the polarization process.%{,  which makes both Reed-Muller and polar codes monomial codes.}

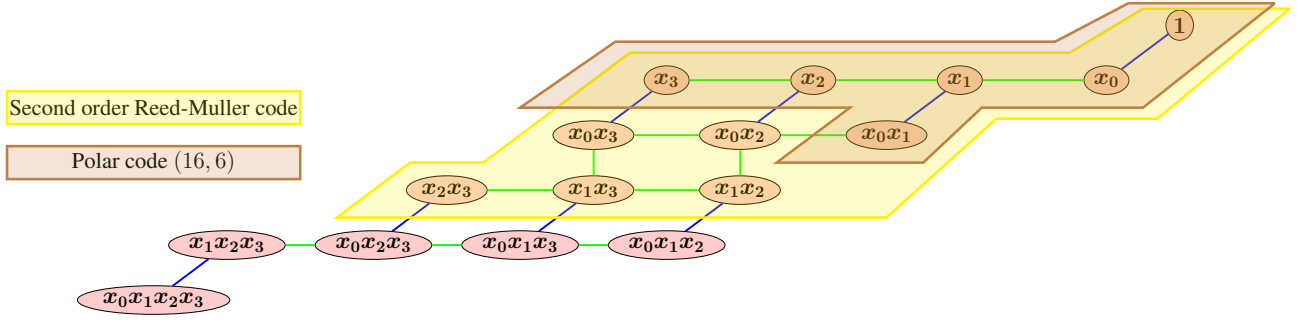
\begin{figure*}[!ht]
\centering
% 1. Use a slightly smaller vertical scale than the 1.6 horizontal scale
% 2. Use 'baseline' to help alignment if needed
\resizebox{0.95\linewidth}{!}{
\begin{tikzpicture}[xscale=1.6, yscale=1.2, % 1.2 is the "sweet spot" for height vs clarity
    state/.style={ellipse, draw, fill=red!20, inner sep=2pt}] % inner sep=2pt is cleaner than array

 % --- NODES (Removed 'array' to eliminate extra vertical padding) ---
 \node[state] at (5,-4) (a) {\Large$\bm{x_0x_1x_2x_3}$};
 \node[state] at (6,-3) (b) {\Large$\bm{x_1x_2x_3}$};
 \node[state] at (8,-3) (c) {\Large$\bm{x_0x_2x_3}$};
 \node[state] at (10,-3) (d1) {\Large$\bm{x_0x_1x_3}$};
 \node[state] at (9,-2) (d2) {\Large$\bm{x_2x_3}$};
 \node[state] at (12,-3) (e1) {\Large$\bm{x_0x_1x_2}$};
 \node[state] at (11,-2) (e2) {\Large$\bm{x_1x_3}$};
 \node[state] at (13,-2) (f1) {\Large$\bm{x_1x_2}$};
 \node[state] at (11,-1) (f2) {\Large$\bm{x_0x_3}$};
 \node[state] at (13,-1) (g1) {\Large$\bm{x_0x_2}$};
 \node[state] at (12,0) (g2) {\Large$\bm{x_3}$};
 \node[state] at (15,-1) (h1) {\Large$\bm{x_0x_1}$};
 \node[state] at (14,0) (h2) {\Large$\bm{x_2}$};
 \node[state] at (16,0) (i) {\Large$\bm{x_1}$};
 \node[state] at (18,0) (j) {\Large$\bm{x_0}$};
 \node[state] at (19,1) (k) {\Large$\bm{1}$};

 % --- EDGES ---
 \draw[very thick, blue] (a) -- (b);
 \draw[very thick, blue] (j) -- (k);
 \draw[very thick, blue] (c) -- (d2);
 \draw[very thick, blue] (d1) -- (e2);
 \draw[very thick, blue] (f2) -- (g2);
 \draw[very thick, blue] (g1) -- (h2);
 \draw[very thick, blue] (h1) -- (i);
 \draw[very thick, blue] (f1) -- (e1);
 \draw[very thick, green] (j) -- (i) -- (h2) -- (g2);
 \draw[very thick, green] (b) -- (c) -- (d1) -- (e1);
 \draw[very thick, green] (d2) -- (e2) -- (f1) -- (g1) -- (h1);
 \draw[very thick, green] (e2) -- (f2) -- (g1) ;

 % --- POLYGONS (Lowered the top y-coords from 1.7 to 1.3 to trim top margin) ---
 \draw [ultra thick, draw=yellow, fill=yellow, fill opacity=0.2](20.5,1.3)--(18.5,1.3)-- (17.5,0.5)--(11.5,0.5)--(9.5,-1.5)--(8.5,-1.5)--(7.5,-2.5)-- (15,-2.5) -- (16.5,-0.7) -- (18.7,-0.7) -- (20.5,1.3);
 \draw [ultra thick, draw=brown, fill=brown, fill opacity=0.2] (20.3,1.4)--(18.3,1.4)-- (17.3,0.7)--(11.3,0.7)--(10,-0.5)--(14.5,-0.5) -- (13.5,-1.5) -- (15.5,-1.5) -- (16.3,-0.5)-- (18.5,-0.5) -- (20.3,1.4);

 % --- LEGEND (Shifted down to sit closer to the diagram) ---
 \node[very thick] at (5,-0.5){\Large Second order Reed-Muller code};
 \draw [ultra thick, draw=yellow, fill=yellow, fill opacity=0.2](3,-0.2) rectangle (7,-0.8);

 \node[very thick] at (5,-1.5){\Large Polar code $(16,6)$};
 \draw [ultra thick, draw=brown, fill=brown, fill opacity=0.2](3,-1.2) rectangle (7,-1.8);

 % --- MANUALLY TIGHTEN BOUNDING BOX ---
 \pgfresetboundingbox
 \path (3, 1.5) rectangle (20.6, -4.2);

\end{tikzpicture}
}
\caption{The Hasse diagram representing $\preceq$ over $\Mon$ for $m=4.$}\label{fig:mon_order_4}
\vspace{-15pt} % Slightly more aggressive caption spacing
\end{figure*}

{ The reliability rule used for polar codes, $\preceq$, was independently discovered by \cite{bardet} and \cite{schurch}, and is called (universal) partial order since it applies to any channel belonging to the class of Binary Discrete Memory-less Channels. Both polar and Reed-Muller codes belong to this family of algebraic codes called \emph{Decreasing Monomial Codes}. A code $\C$ is a decreasing monomial code if it admits a monomial basis $\I\subset\Mon$ satisfying the order relation $\preceq.$ %In our case we will order monomials, hence, the relation will be defined over the set $\Mon.$ 
\begin{definition}\label{def:order}Let $m$ be a positive integer and $f,g\in\Mon.$ When $\deg(f)=\deg(g)=s$ we say that $f\preceq_{\mathbf{sh}} g$ if $\forall\;1\leq\ell\leq s\;\text{ we have }\;  i_\ell \le j_\ell$, where $f=x_{i_1}\dots x_{i_s}$, $g=x_{j_1}\dots x_{j_s}$. 
Define $f\preceq g\quad \text{iff}\quad \exists g^*\in \Mon\;\text{s.t.}\; f\preceq_{\mathbf{sh}} g^*$ and $g^*|g.$
\end{definition}

% Notice that $\preceq$ is indeed a partial order relation since it satisfies the following properties 
% \begin{itemize}
%     \item \emph{reflexive} $f\preceq f$ for any monomial $f$ 
%     \item \emph{transitive} if $f\preceq g$ and $g\preceq h$ then $f\preceq h$ for any monomials $f,g,h$
%     \item \emph{antisymmetric} if $f\preceq g$ and $g\preceq f$ then $f=g$ for any monomials $f,g.$
% \end{itemize}

%The antisymmetry property is due to the fact that 

The notation $f\preceq_{\mathbf{sh}}g$ comes from the fact that one could obtain $g$ from $f$ by positively shifting some of the variables in $f.$ For example $x_0x_1\preceq_{\mathbf{sh}}x_1x_3$ since $x_3$ is a shift by $2$ positions of $x_1$ and $x_1$ is a shift by $1$ position of $x_0$ (see Fig.~\ref{fig:mon_order_4}). Remark that there is a chain relation on the variables, i.e., $x_0\preceq x_1\preceq \dots \preceq x_{m-1}.$ It is a partial order relation, meaning that we might have a pair of monomials that are not comparable, e.g., $x_3x_4$ and $x_1x_5$ are not comparable with respect to $\preceq.$

% \begin{definition}\label{def:order}Let $m$ be a positive integer and $f,g\in\Mon.$ Then $f\weako g$ if and only if $ f|g.$ When $\deg(f)=\deg(g)=s$ we say that $f\preceq_{\mathbf{sh}} g$ if $\forall\;1\leq\ell\leq s\;\text{ we have }\;  i_\ell \le j_\ell$, where $f=x_{i_1}\dots x_{i_s}$, $g=x_{j_1}\dots x_{j_s}$. 
% %
% Define $f\preceq g\quad \text{iff}\quad \exists g^*\in \Mon\;\text{s.t.}\; f\preceq_{\mathbf{sh}} g^*\weako g$.
% \end{definition}

All monomial sets that are used in this tutorial are decreasing monomial sets. 
\begin{definition}
      A set $\I \subseteq \Mon$ is \emph{decreasing}  if and only if ($f \in \I$ and $g \preceq f $) implies $g \in \I$.   
\end{definition}
}

{  We will also consider the ordering of monomials with respect to their degree, i.e., $\I_r=\{f\in\I\mid\deg(f)=r\}.$ Equivalently, $\A_r$ denotes the row indices corresponding to $f$ in $\I_r.$} If $\C(\I)$ is a decreasing monomial code and $r$ is the maximum-degree of $f\in\I$ then the minimum distance of $\C(\I)$ equals $2^{m-r}$ \cite{bardet}.

Note that decreasing monomial codes are not limited to Polar and Reed–Muller codes, but include any code that satisfies the decreasing partial‑order property, even if such codes are not known by specific names.

%Since both Reed-Muller and polar codes admit a maximum-degree monomial set $\I_r=\{f\in\I\mid\deg(f)=r\}$, their minimum distance equals $2^{m-r}.$ We denote $\A_r$ as the corresponding index set to $\I_r$.  

% \vlad{The following well-known polar code will be analyzed in this tutorial. We shall reveal its structure, low-weight codewords and weight distribution. 
% $\CI: (64,32)$ will denote the polar code of length $64$ ($m=6$) and rate $0.5$ I has the following maximum-degree monomials 
% \begin{multline}\label{eq:monom-polar-64}
% \I_3=\{x_0x_2x_5,x_0x_2x_4,x_0x_2x_3,x_0x_1x_5,x_0x_3x_4\\
% x_0x_1x_4,x_0x_1x_3,x_0x_1x_2,x_1x_3x_4,x_1x_2x_4\}.
% \end{multline}
%    %  
% The code has minimum distance $w=2^{6-3}=8.$}

\begin{center}
% \newlength{\mylength}
% %\begin{figure}
% \setlength{\fboxsep}{5pt}
% \setlength{\mylength}{\linewidth}
% \addtolength{\mylength}{-3\fboxsep}
% \addtolength{\mylength}{-3\fboxrule}
% \fbox{
%     \parbox{\mylength}{
%     \setlength{\abovedisplayskip}{0pt}
%     \setlength{\belowdisplayskip}{0pt}
\begin{mdframed}[backgroundcolor=gray!15,linewidth=0pt]
    \textbf{Illustrative Example: Polar Code $\CI:(64,32)$}\\
%We choose the polar code $\CI:(64,32)$\\
%Throughout this section, we use the this example to illustrate the enumeration of various types of codewords. 
For, the code (64,32), we have $m\!=\!6, r\!=\!3$, the minimum distance $w_{\min}\!=\!2^{m-r}\!=\!2^{6-3}\!=\!8$, and the maximum-degree monomials set, $\I_r$ as:
%\begin{multline*}\label{eq:monom-polar-64}
$\I_3=\{x_0x_2x_5,x_0x_2x_4,x_0x_2x_3,x_0x_1x_5,x_0x_3x_4\\
x_0x_1x_4,x_0x_1x_3,x_0x_1x_2,x_1x_2x_3,x_1x_3x_4,x_1x_2x_4\},$ or $\A_3=\{26, 28, 37, 38, 41, 42, 44, 49, 50, 52, 56\}.$
%\end{multline*}
%~~\\
%The minimum distance is $$w_{\min}=2^{m-r}=2^{6-3}=8.$$
% }
% }
\end{mdframed}

\end{center}
%  \caption{Enumeration formulae for codewords of weight $\wm$ (\cite{bardet}) and $1.5\wm$ (this paper) of decreasing monomial codes.} %Formula for $|W_{\wm}|$ is from \cite{bardet} while in this paper we demonstrate the formula for $|W_{1.5\wm}|.$}\label{fig:count-formula}
% \end{figure}

% {\color{red}I have added info on partial order, although I think we don't have space for that.}
% \vlad{Selecting the most reliable channels depends on the channel model. However, universal rules exist, which give partial order relation for the monomials generating the polar code. Decreasing monomial codes are a large family of monomials codes that regroup both polar and Reed-Muller codes. The order relation is rather simple, and on the same time powerful enough to have generated a large number of code properties that we evoke here. 
% \begin{itemize}
%     \item two monomials of same degree $f=x_{i_1}\dots x_{i_r}$ and $g=x_{j_1}\dots x_{j_m-1}$ are ordered $f\preceq_{\mathbf{sh}} g$ 
%     \[{i_1}\leq j_1, \dots, i_{r}\leq j_{r}\]
%     \item two monomials of distinct degrees $f,g$ are ordered $f\preceq_w g$ if $f|g.$
% \end{itemize}
% Combining together these two rules we have 
% $f\preceq g$ either if $f\preceq_{\mathbf{sh}} g$ or if exists $f^{*}$ satisfying $f\preceq_w f^{*}\preceq_{\mathbf{sh}} g.$
% %
% Notice that we have $x_0\preceq x_1\preceq\dots\preceq x_{m-1}.$
% }

%%%%%%%%%%%%%%%%%%%%%%%%%%%%%%%%%%%%%%%%%%%%%%%%%%%%%%%%%%%%%%%%%%%%%%%%
%%%%%%%%%%%%%%%%%%%%%%%%%%%%%%%%%%%%%%%%%%%%%%%%%%%%%%%%%%%%%%%%%%%%%%%%
%{\color{red}Need to change the title of this section.}
\section{Weight-Preserving Transformations}\label{sec:3}

%In this section we shall describe different types of transformations, all preserving the Hamming weight of vectors. Since all binary codes defined in this tutorial are generated by evaluating monomials/polynomials in $m$ variables over the set $\ft^m$, such transformations must necessarily permute the set $\ft^m$ into itself. There are exponentially many such transformations, more precisely, $2^m!$ permutations over the set $\{0,\dots,2^m-1\}.$ As we shall see, not all permutations will be considered. This view looks at the columns of $\bGN$, since these are the evaluation points. However, one can also look at the rows of the matrix, where transformations apply on monomials instead of points. We shall also impose another requirement, namely we shall consider transformations that preserve the affine geometry of the space. These are known as \emph{affine transformations
This section describes weight-preserving transformations. Because our codes evaluate polynomials over $\ft^m$, these transformations must permute $\ft^m$. Instead of considering all $2^m!$ permutations, we restrict our focus to \emph{affine transformations}. These preserve the space's affine geometry and act equivalently on either the columns (evaluation points) or rows (monomials) of $\bGN$.

%}

%Transforming a polynomial $P$ corresponding to a vector $\ev(P)$ produces a new polynomial $Q$ that evaluates to a codeword with the same weight, namely $\wt\ev(P)=\wt\ev(Q)$. For a monomial, the transformation permutes the elements of the row in $F_{2^m}$ that correspond to the original monomial. The transformation of variables can be a translation, defined as $x_i+b_i$ where $b_i \in\{0,1\}$, a linear transformation, defined as $x_i+\sum_{j \in \mathcal{J} \subseteq[0, m-1]\backslash\{i\}} x_j$, or a combination of both, given by $y_i=x_i+$ $\sum_{j \in \mathcal{J} \subseteq[0, m-1] \backslash\{i\}} x_j+b_i$, in $\mathbb{F}_2$.

An affine transformation in $\mathbb{F}_2$ on the $m$ variables input vector $\mathbf{x}=\left(x_0, x_1, \ldots, x_{m-1}\right)$ outputs $\mathbf{y}=\left(y_0, y_1, \ldots, y_{m-1}\right)$ using
%$\mathbf{x}=\left(x_0, x_1, \ldots, x_{m-1}\right) \in \mathbb{F}_2^m$, the transformed variables $\mathbf{y}=\left(y_0, y_1, \ldots, y_{m-1}\right)$ are given by:
\begin{equation}
    \mathbf{y}=\mathbf{A} \mathbf{x}+\mathbf{b},
\end{equation}
%where 
\begin{itemize}
    \item $\mathbf{A}$ is an $m \times m$ invertible matrix over $\mathbb{F}_2$ where $(i, j)$-th entry $A_{i, j}=1$ if $x_j$ appears in the linear combination for $y_i$, and $A_{i, j}=0$ otherwise. This represents the linear transformation 
    \[x_i\rightarrow y_i=\sum_{j \in \mathcal{J} \subseteq[0, m-1]} x_j,\] 
    \item $\mathbf{b}$ is a vector in $\mathbb{F}_2^m$, where $b_i \in\{0,1\}$ represents the translation term for $y_i$, 
    \end{itemize}

 An invertible matrix $\mathbf{A}$ from the general linear group %$\mathrm{GL}(m, 2)$ 
 defines a bijective linear transformation, meaning it maps $\mathbb{F}_2^m$ to itself uniquely (one-to-one and onto). If $\mathbf{A}$ is not invertible, the transformation $\mathbf{x} \mapsto \mathbf{A} \mathbf{x}$ is not bijective; i.e, multiple inputs map to the same output. %The notation $\mathrm{GL}(m, 2)$ is used for general linear transform in $\mathbb{F}_2$ on $m$ variables. %over $\mathbb{F}_2^m$. 
% - Not Onto: The image of the transformation, $\left\{A \mathbf{x} \mid \mathbf{x} \in \mathbb{F}_2^m\right\}$, is a proper subspace of $\mathbb{F}_2^m$, with dimension equal to the rank of $A$, which is less than $m$.
% - Not One-to-One: The kernel of $A, \operatorname{ker}(A)=\{\mathbf{x} \mid A \mathbf{x}=\mathbf{0}\}$, is non-trivial (has dimension $m-\operatorname{rank}(A)>0$ ), so multiple inputs map to the same output.

% In fact, the affine transformation $(\bA, \mathbf{b})$ permutes $D(\bz)$ to $D(\bA \bz+\mathbf{b})$ for $\bz \in \mathbb{F}_2^m$ (see \cite{vlad1.5d} for details).\mohammad{We have removed the def of D(.)}

%%%%%%%%%%%%%%%%%%%%%%%%%%%%%%%%%%%%%%%%%%%%%%%%%%%%%%%%%%%%%%%%%%%%%%%%
\subsection{Permutation automorphism group}
A permutation group is a set of bijective mappings (permutations) of a set (here,  codewords of a code) onto itself, closed under composition and inversion. That is, a permutation $\pi$ is an automorphism of $\C$ if, for any codeword $\bc=\left(c_1, \ldots, c_N\right) \in \C$, we have 
$$
\pi(\boldsymbol{c}) \triangleq\left(c_{\pi(1)}, \ldots, c_{\pi(N)}\right) \in \C.
$$

The permutation automorphisms of $\C$ are automorphisms (structure-preserving bijections) of $\C$ that are, at the same time, permutations of its codewords. Let us emphasize that an automorphism preserves the Hamming distance between codewords and hence the structure of $(\C,\w(\cdot))$. Note that affine transformations act on variables, however, we do not necessarily distinguish between groups of variables transformations and the groups of coordinate permutations that they generate. 
%The automorphism group $\operatorname{Aut}(\C)$ is the subgroup of symmetric group $S_N$ containing all the automorphisms of $C$.

In coding theory, permutation groups often represent symmetries of a code, consisting of coordinate permutations that map codewords to codewords while preserving the code’s structure (its automorphism group). %\vlad{the next details are maybe not necessary}
% The transformations described in the previous section form a group, as they are: 
% \begin{itemize}
%     \item Invertible; Each transformation $\mathbf{y}=A \mathbf{x}+\mathbf{b}$ has an inverse $\mathbf{x}=A^{-1} \mathbf{y}+A^{-1}(-\mathbf{b})=$ $A^{-1} \mathbf{y}+A^{-1} \mathbf{b}$ (since $-\mathbf{b}=\mathbf{b}$ in $\mathbb{F}_2$), which is also affine and in the set.
%     \item Closed under composition. If we take any two of these transformations and compose them (i.e., apply one after the other), the result is another transformation within the same set.
% \end{itemize}

%This section introduces the Lower Triangular Affine (LTA) permutation group \cite{dragoi17thesis} and its extension, the block LTA permutation group, in accessible terms. Through illustrative examples, we demonstrate their transformations and applications in automorphism ensemble decoding. 

The permutations group of a Reed-Muller code  in $m$ variables includes all coordinate permutations induced by the full affine group.  % acting on $\ft^m.$}
 Indeed, any affine transformation applied to monomials of degree less than of equal to $r$ will give polynomials satisfying the same degree condition. If the linear transformation matrix $\mathbf{A}$ is a lower triangular matrix having $A_{i,i}=1$ then we will have the lower triangular affine (LTA) group. $\text{LTA}(m, 2)$ is a group that globally leaves invariant any decreasing monomial code, hence, implicitly polar codes and Reed-Muller codes \cite{bardet}.  %An element $(\mathbf{A}, \mathbf{b}) \in \text{LTA}(m, 2)$ consists of a lower triangular matrix $\mathbf{B} \in \text{GL}(m, 2)$ with 1s on the diagonal and a translation vector $\varepsilon \in \mathbb{F}_2^m$. 
The LTA acts on a monomial $f = \prod_{i \in \text{ind}(f)} x_i$ by transforming each $x_i$ into:
\begin{equation}\label{eq:linear_form}
y_i = x_i + \sum_{j<i} A_{i,j} x_j + b_i.\vspace{-8pt}
\end{equation}
{
\begin{equation}\label{eq:action-monomial}
(\bA,\bb)\cdot f=\prod\limits_{i\in\ind(f)}\left(x_i + \sum_{j<i} A_{i,j} x_j + b_i\right).
\end{equation}
}
For $f = x_0 x_1$, with $\mathbf{A} = \begin{pmatrix} 1 & 0 \\ 1 & 1 \end{pmatrix}$, $\mathbf{b} = \begin{pmatrix}0\\1\end{pmatrix}$, we get $y_0 = x_0$, $y_1 = x_1 + x_0 + 1$. We denote this action by
\[
(\mathbf{A}, \mathbf{b}) \cdot f = x_0 (x_1 + x_0 + 1). %\cite{bardet2016}.
\]
%If we consider all LTA transformations $(\mathbf{A}, \mathbf{b}) \in \text{LTA}(m, 2)$, the group action on $f$ is denoted by $$\text{LTA}(m, 2) \cdot f.$$ 
In matrix terms, $\text{LTA}(m, 2)$ permutes columns of $\bGN$ or evaluation points, preserving the code $\mathcal{C}(\mathcal{I})$. Example \ref{ex:transforms} demonstrates LTA on $f=x_0x_1$.

\begin{example}\label{ex:transforms}
For $m=2$, transformations of $x_0 x_1$ are tabulated in the following table.

\begin{table}[h!]\vspace{-10pt}
\centering
\renewcommand{\arraystretch}{1.2}
\begin{tabular}{|c|c|c|}
\hline\rowcolor{c2}
\textbf{Transformation} & \textbf{Evaluation} & \textbf{Codeword} \\
\hline
\makecell[l]{Original: $x_0x_1$} & 
\makecell[l]{$(0,0)\to 0$ \\ $(0,1)\to 0$ \\ $(1,0)\to 0$ \\ $(1,1)\to 1$} & 
$[1\ 0\ 0\ 0]$ \\
\hline
\makecell[l]{$y_0 = x_0+1, y_1 = x_1$ \\ $\to y_0y_1=(x_0+1)x_1$} & 
\makecell[l]{$(0,0)\to 0$ \\ $(0,1)\to 1$ \\ $(1,0)\to 0$ \\ $(1,1)\to 0$} & 
$[0\ 0\ 1\ 0]$ \\
\hline
\makecell[l]{$y_0 = x_0, y_1 = x_1+1$ \\ $\to y_0y_1=x_0(x_1+1)$} & 
\makecell[l]{$(0,0)\to 0$ \\ $(0,1)\to 0$ \\ $(1,0)\to 1$ \\ $(1,1)\to 0$} & 
$[0\ 1\ 0\ 0]$ \\
\hline
\makecell[l]{$y_0 = x_0+1, y_1 = x_1+1$ \\ $\to y_0y_1=(x_0+1)(x_1+1)$} & 
\makecell[l]{$(0,0)\to 1$ \\ $(0,1)\to 0$ \\ $(1,0)\to 0$ \\ $(1,1)\to 0$} & 
$[0\ 0\ 0\ 1]$ \\
\hline
\end{tabular}\vspace{-10pt}
\end{table}

% Observe that the resulting codewords have the same weight. Note that other transformations exist, such as the following, which evaluate to one of the codewords above.\\
% Transformation: $y_0=x_0+1, y_1=x_1+x_0+1 \rightarrow \left(x_0+1\right)\left(x_1+x_0+1\right)$.\\
%     Evaluation: $(0,0) \rightarrow 1 \cdot 1=1,(0,1) \rightarrow 1 \cdot 0=0,(1,0) \rightarrow 0 \cdot 0=0,(1,1) \rightarrow 0 \cdot 1=0$.\\
%     Codeword: $[0\;0\;0\;1]$.
%To avoid this, the 

% \begin{tikzpicture}[
%   grow=down,
%   level distance=2cm,
%   sibling distance=3cm,
%   every node/.style={%draw, rectangle, rounded corners,
%   align=center, minimum height=1.2cm, minimum width=3cm},
%   edge from parent/.style={draw, -latex},
%   label/.style={font=\small, above, pos=0.5}
% ]

% % Root node
% \node {$\begin{array}{c} x_0 x_1 \\ [1\;0\;0\;0] \end{array}$}
%   % Children
%   child {
%     node {$\begin{array}{c} (x_0 + 1)x_1 \\ [0\;0\;1\;0] \end{array}$}
%     edge from parent node[label] {$x_0 \to x_0 + 1, x_1 \to x_1$}
%   }
%   child {
%     node {$\begin{array}{c} x_0 (x_1 + 1) \\ [0\;1\;0\;0] \end{array}$}
%     edge from parent node[label] {$x_0 \to x_0, x_1 \to x_1 + 1$}
%   }
%   child {
%     node {$\begin{array}{c} (x_0 + 1)(x_1 + 1) \\ [0\;0\;0\;1] \end{array}$}
%     edge from parent node[label] {$x_0 \to x_0 + 1, x_1 \to x_1 + 1$}
%   };

% \end{tikzpicture}

The matrix $\mathbf{A}$ for all transformations is
    \[
        \mathbf{A}=\left(\begin{array}{ll}
    1 & 0 \\
    0 & 1
    \end{array}\right),\vspace{-5pt}
    \]
    and the translation vectors are
    \[
    \mathbf{b}\in\left\{\binom{0}{0}, \binom{1}{0}, \binom{0}{1}, \binom{1}{1} \right\}.\vspace{-3pt}
    \]
    %For instance, 
    $$
%\left(x_0, x_1\right) \rightarrow
\left[\begin{array}{ll}
1 & 0 \\
0 & 1
\end{array}\right]\left[\begin{array}{l}
x_0 \\
x_1
\end{array}\right]+\left[\begin{array}{l}
b_0 \\
b_1
\end{array}\right]=\left(x_0+b_0, x_1+b_1\right)
$$
\end{example}

\subsection{Monomial Orbit and Minimum-weight Codewords}\label{ssec:orbit_wm}
%This section examines the action of the Lower Triangular Affine (LTA) transform on monomials \cite{dragoi17thesis}, elucidating its role in shaping the minimum-weight codewords. Visual interpretations of these actions, for example, Young diagrams, will be used to express minimum-weight codewords counting methods. We further demonstrate the equivalent representation of cosets formed by transforming the row indices of the polar transform and decomposition of minimum-weight codewords in terms of row combination \cite{rowshan2023formation,rowshan2023minimum}. 

%\vlad{page 5, col 2, line 2-8: The definition of orbit conflates the general definition with the special case of LTA.  Please define general case and then specialize to LTA.}
The term ``orbit" comes from group theory. Given a group $\mathsf{G}$ that acts on a set $\mathcal{X}$, the orbit of $x\in\mathcal{X}$ with respect to $\mathsf{G}$ is the set of images, or elements where $x$ can travel by means of $\mathsf{G}$, i.e., $\{g\cdot x\mid g\in \mathsf{G}\}.$ In our case, $x$ is a monomial, the starting point in the ``space" that ``moves" through a set of related polynomials under the action of LTA, similar to a planet’s path around a star, visiting a predictable path of positions. %The orbit is the complete set of polynomials you can reach by applying all possible LTA transformations to the starting monomial.}
%The term ``orbit" comes from {group theory based on} the idea of a monomial as a starting point in the ``space" ``moving" through a set of related monomials under the action of the LTA permutation group, similar to a planet’s path around a star, visiting a predictable path of positions. The orbit is the complete set of polynomials you can reach by applying all possible LTA transformations to the starting monomial. %Imagine a monomial, like $x_2$, as a starting point in the "space" of all possible monomials for a polar code with $m$ variables. The LTA group provides a set of "moves" (transformations) that change the monomial into other monomials by rearranging or combining variables in a structured way. %It's called an orbit because these transformations trace out a connected set of points (monomials) that are equivalent in terms of the code's structure.
Formally, the $\text{LTA}(m, 2)$ action on a monomial $f$ forms an orbit $$\text{LTA}(m, 2) \cdot f = \{(\mathbf{A}, \mathbf{b}) \cdot f \mid (\mathbf{A}, \mathbf{b}) \in \text{LTA}(m, 2)\}.$$
Next step would be to determine $|\Alow\cdot f|$. Using $|\Alow|$ overestimates this number. For example, let $f=x_0x_1.$ Two distinct transformations map $f$ into the same output 
\begin{align*}
\left(\begin{pmatrix} 1 & 0 \\ 1 & 1 \end{pmatrix},\begin{pmatrix}  0 \\ 0 \end{pmatrix}\right)\cdot f &= x_0(x_1+x_0)=x_0x_1+x_0\\
\left(\begin{pmatrix} 1 & 0 \\ 0 & 1 \end{pmatrix},\begin{pmatrix}  0 \\ 1 \end{pmatrix}\right)\cdot f &= x_0(x_1+1)=x_0x_1+x_0.
\end{align*}

Hence, a subgroup of $\Alow$ was proposed ($\Alow_f$) in order to better characterize the orbits. This subgroup depends on the variables in $f$ and restricts $\Ab$ in a simple manner, i.e., all rows and columns corresponding to variables not in $f$ are set to zero, except the diagonal of $\bA$ where $A_{i,i}=1.$ Equation \eqref{eq:action-monomial} becomes 
{\vspace{-10pt}
\begin{equation}\label{eq:LTAaction-monomial}
(\bA,\bb)\cdot f=\prod\limits_{i\in\ind(f)}\left(x_i + \sum_{j<i,j\notin \ind(f)} A_{i,j} x_j + b_i\right).
\end{equation}
}Now, one can estimate the number of elements in $\Alow_f$, as follows.\\
%\begin{itemize}
 %   \item 
 1) LTA matrices $\mathbf{A}$, satisfying $A_{i, j}=0,j<i,j\in\ind(f),i\notin\ind(f)$. %In other words all rows of $\Ab$ indexed by variables that are not in $f$, and all column of $\Ab$ indexed by variables in $f$ are set to zero, except for the diagonal. 
    To estimate the number of such matrices, denote $J_f(i) = \{j < i \mid j \notin \text{ind}(f)\}.$ Hence, we map a monomial $f=x_{i_0}\dots x_{i_{r-1}}$ to an integer partition $\lambda_f=(|J_f(i_{r-1})|,\dots |J_f(i_0)|)$ inside the $\deg(f)\times (m-\deg(f)$ grid (\ref{ex:young-diagram1}). The rows of this grid are indexed by the variables in $f$, while the columns by the variables in $\check{f}$. For each row (say variable $x_{i_l}$), we highlight the boxes in this grid (in blue) that correspond to column variables which are smaller than $x_{i_l}$.} We have $\lambda_f=(i_r-r+1,\dots,i_1-1, i_0-0).$ The total number of free variables, or free entries in $\bA$ equals %{\color{red}We need to have a general definition for $|\lambda_f(g)|$ or $|\lambda_g(f)|$: For $g=x_{j_{0}}\dots x_{j_{l-1}}$ satisfying $g|f$, we have $\lambda_f(g)$ is the partition of length $l$ defined by $\lambda_f(g)=(|J_f(j_{l-1})|,\dots |J_f(j_0)|)$, which yields 
% \begin{equation}\label{eq:lambda_f_g}
%     |\lambda_f(g)|=\sum_{i\in\ind(g)}|J_f(i)|.
% \end{equation}}
   \begin{equation}2^{|\lambda_f|}, \text{ where }
         |\lambda_f| = \sum_{i \in \text{ind}(f)} |J_f(i)|.
     \end{equation}
     In general, for $g=x_{j_{0}}\dots x_{j_{l-1}}$ satisfying $g|f$, we define length-$l$ partition $\lambda_f(g)=(|J_f(j_{l-1})|,\dots |J_f(j_0)|)$, which yields 
\begin{equation}\label{eq:lambda_f_g}
    |\lambda_f(g)|=\sum_{i\in\ind(g)}|J_f(i)|.
\end{equation} 
 The power of two arises because each entry $A_{i, j}\in\{0,1\}$, except the fixed ones.\\% can independently take on one of two possible values: 0 or 1. 
    % The total number of LT matrices $\mathbf{A}$ is $2^{\lambda_f}$. The power of two arises because each entry $A_{i, j}$ can independently take on one of two possible values: 0 or 1. 
    % \vlad{Each monomial maps into a partition using $J_f(i)$, give example.}
    % For this purpose, we introduce the parameter $\lambda_f$ as
    % \begin{equation}
    %     \lambda_f = \sum_{i \in \text{ind}(f)} |J_f(i)|,
    % \end{equation}
    % where \[ J_f(i) = \{j < i \mid j \notin \text{ind}(f)\}. \]
    % This accounts for the number of free entries $A_{i, j} \in\{0,1\}$ in row $i\in\ind(f)$, which directly corresponds to the number of free variables in each linear form $y_i$ in \eqref{eq:linear_form}. 
 %   \item 
 2) Translation vectors $\mathbf{b}$, satisfy $b_i=0,i\notin\ind(f)$, and there are $2^{\deg(f)}$, as $\forall j\in\ind(f),b_j\in\{0, 1\}$. 
%\end{itemize}
Since $\mathbf{A}$ and $\mathbf{b}$ are independent, the total count is obtained by simply multiplying the two numbers%their individual counts according to combinatorial principles. Therefore, the cardinality of the orbit is given by:
{
\begin{equation}
|\Alow \cdot f| = |\Alow_f\cdot f| = 2^{\text{deg}(f) + |\lambda_f|}.
\end{equation}
}
%where $\lambda_f = \sum_{i \in \text{ind}(f)} |J_f(i)|$, and $J_f(i) = \{j < i \mid j \notin \text{ind}(f)\}$ counts free variables \cite{bardet}. 

%\vlad{Maybe add Young diagram here}

%One way of visualizing the action of $\Ab$ is using \emph{Young diagrams}. Indeed, any monomial maps into $\lambda_f$ which is an integer partition inside the $\deg(f)\times (m-\deg(f)$ grid \cite{dragoi17thesis} (see Example \ref{ex:young-diagram1}).
\begin{example}\label{ex:young-diagram1}
    Let $m=7$ and $f=x_1x_3x_5\Rightarrow\lambda_f=(5-2,3-1,1-0)=(3,2,1).$ Thus, $|\text{LTA}(m, 2) \cdot f| = 2^{\text{deg}(f) + |\lambda_f|}=2^{3+6}$.      
    %A Young diagram for this example on a $3\times 3$ grid is represented in Fig. \ref{fig:Young-diagram}. The rows are labeled by the variables in $f$, and columns by those in $\widecheck{f}$. Each blue cell indicates a free variable in the columns used to form linear combinations with the row variables of $f$. 
   %\begin{figure}[h]
    \begin{center}
     \resizebox{0.12\textwidth}{!}{
    \begin{tikzpicture}[inner sep=0in,outer sep=5in]
      \node(m) {
\begin{ytableau}
 *(blue! 40) *  &  0 & 0 \\
 *(blue! 40) *  & *(blue! 40) * & 0 \\
*(blue! 40) *  &*(blue! 40) *   &*(blue! 40) *  \\
\end{ytableau}};
\node at (-0.65,1.25) {$x_0$};
\node at (0,1.25) {$x_2$};
\node at (0.6,1.25) {$x_4$};
\node at (-1.35,0.6) {$x_1$};
\node at (-1.35,-0.05) {$x_3$};
\node at (-1.35,-0.7) {$x_5$};
\end{tikzpicture}
}
 \end{center}
    %   \caption{Young diagram corresponding to $f=x_1x_3x_5.$ Rows represents variables in $f$ while columns variables in $\widecheck{f}.$ $\lambda_f=(3,2,1)$ - $x_1$ can be combined only with $x_0$, $x_3$ with $x_0,x_2$ and $x_5$ with $x_0,x_2,x_4.$   }
    %   \label{fig:Young-diagram}
   %\end{figure}

The Young diagram corresponding to $f=x_1x_3x_5$ is illustrated above. Rows represent variables in $f$ while columns represent variables in $\widecheck{f}.$ The variable $\lambda_f=(3,2,1)$ means $x_1$ can be combined only with $x_0$; $x_3$ with $x_0,x_2$; and $x_5$ with $x_0,x_2,x_4.$ 
\end{example}

\begin{example}\label{ex:orbit_x1}
    Let $m=3, f=x_1.$ %, find the orbit of this monomial, $\operatorname{Orb}(f)$. 
    %
    %To find the orbit, we need to consider all possible LTA permutations within the group permutation $\Alow$. We know that any transformation $\mathbf{A} \mathbf{x}+\mathbf{b}$ could keep $x_1$ unchanged or introduce linear combination involving $x_0$,  with/without translation.
    The orbit of $f$ is
$$
\Alow\cdot (x_1)=\{x_1+a_{1,0}x_0+b_1 \mid a_{1,0},b_1\in\{0,1\}\}.
$$
{Fig. \ref{fig:orbit-polynomials} shows all polynomials in this orbit.} For example, 
\begin{itemize}
\item With $\mathbf{A}=\left[\begin{array}{lll}1 & 0 & 0 \\ 0 & 1 & 0 \\ 0 & 0 & 1\end{array}\right]$ and $\mathbf{b}=(0,1,0)$, we get $f^{\prime}=x_1+1$.
\item With $\mathbf{A}=\left[\begin{array}{lll}1 & 0 & 0 \\ 1 & 1 & 0 \\ 0 & 0 & 1\end{array}\right]$ and $\mathbf{b}=(0,1,0)$, we get $f^{\prime}=x_1+x_0+1$.
%\item With $\mathbf{A}=\left[\begin{array}{lll}1 & 0 & 0 \\ 1 & 1 & 0 \\ 0 & 0 & 1\end{array}\right]$ and $\mathbf{b}=(0,0,0)$, we get $f^{\prime}=x_1+x_0$.
\end{itemize}
% In general, the orbit of $f=x_1$ is
% $$
% \operatorname{Orb}(x_1)=\{x_1+a_{1,0}x_0+b_1 \mid a_{1,0},b_1\in\{0,1\}\}.
% $$
Observe that $|\Alow\cdot(x_1)|=2^2$ because we have two variables $a_{1,0},b_1\in\{0,1\}$; that is, $2\times2\!=\!2^2$. 
% The codewords associated to the polynomials in $\Alow_{x_1}\cdot x_1$ are 
% \begin{center}
% \resizebox{0.4\textwidth}{!}{    
% \begin{tabular}{|l|c|}
% \hline Polynomials $P$ & Binary Vectors $\ev(P)$ \\
% \hline $x_1$ & $[1\;1\;0\;0\;1\;1\;0\;0]$ \\
% \hline $x_1+1$ & $[0\;0\;1\;1\;0\;0\;1\;1]$ \\
% \hline $x_1+x_0$ & $[0\;1\;1\;0\;0\;1\;1\;0]$ \\
% \hline $x_1+x_0+1$ & $[1\;0\;0\;1\;1\;0\;0\;1]$ \\
% \hline
% \end{tabular}
% }
% \end{center}
%Thus, the orbit of $x_1$ includes
% $$
% \Alow_{x_1}\cdot x_1=\left\{x_1, x_1+1, x_1+x_0, x_1+x_0+1\right\}.
% $$

\begin{figure}[ht]
    \centering
    \resizebox{0.25\textwidth}{!}{
    \begin{tikzpicture}
        \newcommand{\co}{(0.75,5) circle (4)}
        \node[circle, thick, draw=brown, minimum size=40pt] (c1) at (-1.5,2.5){$\bm{~~\Alow\cdot x_1~~}$};
        \foreach \i in {0,90,180,270} {
        \fill[black] (c1.\i) circle [radius=2pt]{};
        %\node at ($(c1.\i) + (0.5,0)$) {$P_1$}; % Place $P_1$ outside each circle
        \node at ($(c1.0) + (0.5,0)$) {$x_1$};
        \node at ($(c1.90) + (0,0.5)$) {$x_1+1$};
        \node at ($(c1.180) + (-1,0)$) {$x_1+x_0$};
        \node at ($(c1.270) + (0,-0.5)$) {$x_1+x_0+1$};
    }
    \end{tikzpicture}
    }
    \caption{Visualization of %the orbit of monomial $f=x_1$, 
    $\Alow_{x_1}\cdot x_1.$}
    \label{fig:orbit-polynomials}\vspace{-10pt}
\end{figure}
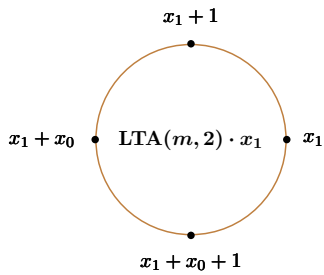

%In summary, we have 
%\begin{table}\centering

%\end{table}
%where the weight of all vectors/codewords are the same, but the elements have been permuted.
\end{example}

%\vlad{Mention that minimum distance of polar/RM code is $\wm=2^{m-r}.$}

Since the transformation preserves the weight of codewords, the minimum-weight codewords lie in such orbits for $f \in \mathcal{I}_r$ \cite{bardet}. Hence, the total number of minimum-weight codewords is the sum of the cardinality of all such orbits; that is,
\begin{equation}\label{eq:sum_A_wm}
    A_{2^{m-r}}(\I)=\sum\limits_{f\in \I_r}|\text{LTA}(m, 2) \cdot f|=\sum\limits_{f\in \I_r}2^{\deg(f)+|\lambda_f|}.
\end{equation}

%W8 = 920, W12 = 25472, W14 = 32768.

\begin{center}
%\newlength{\mylength}
%\begin{figure}
% \setlength{\fboxsep}{5pt}
% \setlength{\mylength}{\linewidth}
% \addtolength{\mylength}{-3\fboxsep}
% \addtolength{\mylength}{-3\fboxrule}
% \shadowbox{
%     \parbox{\mylength}{
%     \setlength{\abovedisplayskip}{0pt}
%     \setlength{\belowdisplayskip}{0pt}
\begin{mdframed}[backgroundcolor=gray!10,linewidth=0pt]
    Minimum weight codewords of $\CI:(64,32)$
    \centering
        \footnotesize
 \begin{tabular}{c|c|c}
        \toprule
        $f\in\I_r$ &  $\lambda_f$ & $|\Alow_f\cdot f|=A_{\wm}(f)$\\
\midrule
        $x_0x_2x_5$ & $(3,1,0)$&$2^{3+4}=128$\\
        $x_0x_2x_4$ & $(2,1,0)$&$2^{3+3}=64$\\
        $x_0x_2x_3$ & $(1,1,0)$&$2^{3+2}=32$\\
        $x_0x_1x_5$ & $(3,0,0)$&$2^{3+3}=64$\\
        $x_0x_1x_4$ & $(2,0,0)$&$2^{3+2}=32$\\
        $x_0x_1x_3$ & $(1,0,0)$&$2^{3+1}=16$\\
        $x_0x_1x_2$ & $(0,0,0)$&$2^{3}=8$\\
        $x_1x_3x_4$ & $(2,2,1)$&$2^{3+5}=256$\\
        $x_1x_2x_4$ & $(2,1,1)$&$2^{3+4}=128$\\
        $x_1x_2x_3$ & $(1,1,1)$&$2^{3+3}=64$\\
         $x_0x_3x_4$ & $(2,2,0)$&$2^{3+4}=128$\\
         \midrule
         \multicolumn{3}{c}{Total: $A_{8}(\I)=\sum_{f\in\I_r} |\Alow_f\cdot f| =920.$}\\
         \bottomrule
         \multicolumn{3}{p{0.9\textwidth}}{A program for enumerating arbitrary polar codes, including our example, is available on GitHub; see \cite{Rowshan2025weightGITHUB}.}
    \end{tabular}
%}
%} 
%\mohammad{Here, we have 11 elements in $\I_3$ but the previous frame/box (working example), there are 10 elements. }

\end{mdframed}
    
\end{center}
% \begin{table}[h]
%     \centering
%     \begin{tabular}{c|c|c}
%         \toprule
%         $f$ &  $\lambda_f$ & $\Alow_f\cdot f$\\
% \midrule
%         $x_0x_2x_5$ & $(3,1,0)$&$2^{3+4}=128$\\
%         $x_0x_2x_4$ & $(2,1,0)$&$2^{3+3}=64$\\
%         $x_0x_2x_3$ & $(1,1,0)$&$2^{3+2}=32$\\
%         $x_0x_1x_5$ & $(3,0,0)$&$2^{3+3}=64$\\
%         $x_0x_1x_4$ & $(2,0,0)$&$2^{3+2}=32$\\
%         $x_0x_1x_3$ & $(1,0,0)$&$2^{3+1}=16$\\
%         $x_0x_1x_2$ & $(0,0,0)$&$2^{3}=8$\\
%         $x_1x_3x_4$ & $(2,2,1)$&$2^{3+5}=256$\\
%         $x_1x_2x_4$ & $(2,1,1)$&$2^{3+4}=128$\\
%         $x_1x_2x_3$ & $(1,1,1)$&$2^{3+3}=64$\\
%          $x_0x_3x_4$ & $(2,2,0)$&$2^{3+4}=128$\\
%          \midrule
%          \multicolumn{3}{c}{Total $A_{8}(\I)=920.$}\\
%          \bottomrule
%     \end{tabular}
%     \caption{Number of minimum-weight codewords of the polar code $\CI :(64,32)$}
%     \label{tab:min-polar-64}
% \end{table}

\textbf{Row combinations.}
%{\color{red}Need to define set M and $g_r$ and review this part, the polar transform is $\mathbf{G}_{2^m}.$} 
Denote $\S_i=\supp(\bin(i))$ and suppose monomial $f$ corresponds to row index $i$ of the polar transform {$\mathbf{G}_{2^m}$} with binary representation $\bin(i)$, then the corresponding orbit $\cO_i$ in terms of {$\mathbf{G}_{2^m}$}-rows is the collection of all subsets of rows that sum to a codeword of weight $2^{\w(\bin(i))}$ \cite{rowshan2023formation}. That is,
\begin{equation}\label{eq:vector_equality_all}
    \cO_i=\bigcup_{\J\subseteq\K_i}\cO_i(\J),
\end{equation}
%\vlad{The question $\I \setminus[0,i]$ refers to the fact that $\I $ is the notation we have used for monomials and $\A $ the notation for integer indices.}
where the set of row indices corresponding to $\J$ is
\begin{equation}\label{eq:vector_equality}
    \cO_i(\J)=\{i\}\cup\J\cup\M(\J), 
\end{equation}
{and} the set $\Ki\supseteq\J$ is defined as %\vlad{the $j \in \I\backslash[0,i]$ ???}
\cite{rowshan2023formation}
%a set of $G_{2^m}$ row indices, called core rows, and denoted by $\{i\}\cup\K_i$ \cite{rowshan2023formation}. %For $f = x_0 x_1 x_2$, the orbit maps to a set of row 0, with weight $2^{3-3} = 1$. 
\begin{equation}\label{eq:Ki}
    \K_{i} = \{j \in \A\backslash[0,i]\colon |\S_j\backslash\S_i|=1\}. 
    %\{| i_j\in\{0,1\} \text{ if } i_j=0, i_z\in\{0,1\} \text{ if } i_j=0 \text{ and } z>j \text{ and } z\not\in\supp(i_{m-1}\cdots i_0)\}
    %\{j \in [i+1,N-1]\colon |\Sj\setminus\Si|=1, |\Sj|=|\Si| \text{ or }|\Si|+1 \}
\end{equation}
%For definition of set $\M(\J)$, the readers are referred to \cite[$\M$-Contruction]{rowshan2023formation}. 
%%
{
The set $\M(\J)$ is uniquely determined by 
{\footnotesize\begin{equation*}\label{eq:M_def}
\M(\J)
\!\triangleq\!
\Bigl\{
m\!\in\!\A\setminus[0,i]:
\bigl|\{\J'\!\subseteq\!\J:\ |\J'|\!\ge\!2,\ m_{\J'}\!=\!m\}\bigr|
\text{ is odd}
\Bigr\}.
\end{equation*}}
For each $\J'\subseteq\J$ with $|\J'|\ge2$ and
$\{\S_j\setminus\S_i:j\in\J'\}$ pairwise disjoint,
the support of index $m_{\J'}$ is defined by 
\begin{equation*}\label{eq:mJ_def}
\S_{m_{\J'}} =
\Bigl(\bigcup_{j\in\J'}(\S_j\setminus S_i)\Bigr)
\cup
\Bigl(\S_i\cap\!\!\bigcap_{j\in\J'}\S_j\Bigr).
\end{equation*}
}

\begin{figure*}[ht]
\centering
\resizebox{0.8\textwidth}{!}{
\begin{tikzpicture}
% --- Left Side (1.5wm) ---
\draw (-5,1.5) node {$\ev(x_2x_3x_4)$}
      (-3,0.5) node {$\ev(x_2x_5x_6)$}
      (1.5,1) node {\small$\ev(x_2x_3x_4x_5x_6)$};

% Top Rectangle Left
\draw (-4,1.7) rectangle (-1.2,1.3);
\pattern[pattern=north west lines, pattern color=green] (-4,1.7) --(-4,1.3) --(-1.9,1.3)--(-1.9,1.7)--cycle;
\draw[dashed] (-1.9,1.1) -- (-1.9,1.9);
\pattern[pattern=crosshatch, pattern color=red] (-1.9,1.3) --(-1.2,1.3) --(-1.2,1.7)--(-1.9,1.7)--cycle;
\draw[latex-latex] (-1.9,2.0) -- node[above, yshift=-2pt] {\footnotesize$\frac{1}{4}2^{7-3}$} (-1.2,2.0);
\draw[latex-latex] (-4,2.0) -- node[above, yshift=-2pt] {\footnotesize$\frac{3}{4}2^{7-3}+$} (-1.2,2.0);

% Bottom Rectangle Left
\draw (-1.9,0.3) rectangle (-1.9+2.8,0.7);
\pattern[pattern=north west lines, pattern color=green] (-1.2,0.3) --(-1.2,0.7) --(0.9,0.7)--(0.9,0.3)--cycle;
\draw[dashed] (-1.2,0.1) -- (-1.2,0.9);
\pattern[pattern=crosshatch, pattern color=red] (-1.2,0.7) --(-1.9,0.7) --(-1.9,0.3)--(-1.2,0.3)--cycle;

\draw[-{Stealth[length=5pt]},thick] (-1.6,1.5) to [bend right=15] (0,1);
\draw[-{Stealth[length=5pt]},thick] (-1.6,0.5) to [bend left=15] (0,1);

\draw[latex-latex] (-4,0.0) -- node[below, yshift=2pt] {\footnotesize$1.5\times 2^{7-3}$} (0.9,0.0);
\draw (-2,-0.6) node {\bf a) $1.5\wm$};

% --- Right Side (1.75wm) ---
\draw (-5+10,1.5) node {$\ev(x_2x_3x_4)$}
      (-3+9.5,0.5) node {$\ev((x_2+x_0)x_5x_6)$}
      (1.5+10.5,1) node {\small$\ev((x_0+1)x_2x_3x_4x_5x_6)$};

% Top Rectangle Right
\draw (-4+10,1.7) rectangle (-1.2+10,1.3);
\pattern[pattern=north west lines, pattern color=green] (6,1.7) --(6,1.3) --(8.35,1.3)--(8.35,1.7)--cycle;
\draw[dashed] (8.35,1.1) -- (8.35,1.9);
\pattern[pattern=crosshatch, pattern color=red] (8.35,1.3) --(8.8,1.3) --(8.8,1.7)--(8.35,1.7)--cycle;
\draw[latex-latex] (8.35,2.0) -- node[above, yshift=-2pt] {\footnotesize$\frac{1}{8}2^{7-3}$} (8.8,2.0);
\draw[latex-latex] (6,2.0) -- node[above, yshift=-2pt] {\footnotesize$\frac{7}{8}2^{7-3}+$} (8.8,2.0);

% Bottom Rectangle Right
\draw (8.35,0.3) rectangle (10.9,0.7);
\pattern[pattern=north west lines, pattern color=green] (8.8,0.3) --(8.8,0.7) --(10.9,0.7)--(10.9,0.3)--cycle;
\draw[dashed] (8.8,0.1) -- (8.8,0.9);
\pattern[pattern=crosshatch, pattern color=red] (8.8,0.7) --(8.35,0.7) --(8.35,0.3)--(8.8,0.3)--cycle;

\draw[-{Stealth[length=5pt]},thick] (8.5,1.5) to [bend right=15] (10,1);
\draw[-{Stealth[length=5pt]},thick] (8.5,0.5) to [bend left=15] (10,1);

\draw[latex-latex] (6,0.0) -- node[below, yshift=2pt] {\footnotesize$1.75\times 2^{7-3}$} (10.9,0.0);
\draw (8,-0.6) node {\bf b) $1.75\wm$};
\end{tikzpicture}
}
\caption{Two distinct weights in the sum of two orbits $\Alow\cdot x_2x_3x_4+\Alow\cdot x_2x_5x_6$, for $m=7.$}\label{fig:example1_intro}
\end{figure*}
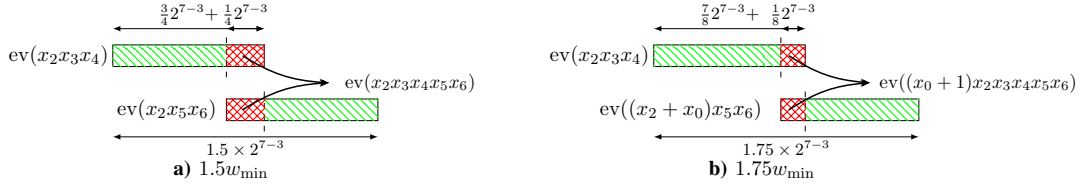

Therefore, the minimum-weight codewords $\bc_{w_{\min}}$ are formed {by summing the rows $\bg_r$ of $\bGN$ with indices $r\in\cO_i(\J)$ where $i\in\A_r,\J\subseteq\Ki$ as
\begin{equation}\label{eq;row-sum}
    \bc_{w_{\min}} = \sum_{r\in\cO_i(\J)} \bg_r.
\end{equation}

\begin{example}[Equivalent to Example \ref{ex:orbit_x1}]\label{ex:orbit_i=1}
For $m=3$ and $i=5=(101)_2$, which corresponds to $f=x_1$, according to \eqref{eq:Ki}, we have the row set $\K_i=\{6,7\}=\{(110)_2,(111)_2\}$. Here, we get $\MJ=\emptyset$ for every $\J\subseteq\K_i$. % Hence, we have
%the orbit of $x_0 x_1$ has weight $2^{3-2} = 2$, corresponding to a set of $G_8$ rows. %A Young diagram for $f = x_0 x_1 x_2$ shows $\lambda_f = (2, 1, 0)$, visualizing free variables as boxes \cite{dragoi2016}. 
{Thus, according to \eqref{eq:vector_equality} and \eqref{eq;row-sum}, the row summation $\bg_{5}+\sum_{j\in\J}\bg_j$ for every subset $\J\subseteq\{6,7\}$ form a codeword of weight $2^{2}=4$. Fig.~\ref{fig:row-combin} lists them.}

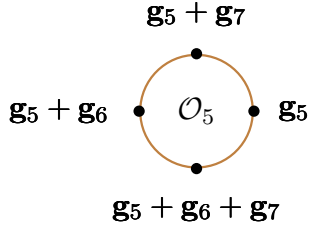
\begin{figure}[!h]\vspace{-5pt}
    \centering
    \resizebox{0.25\textwidth}{!}{
    \begin{tikzpicture}
        \newcommand{\co}{(0.75,5) circle (4)}
        \node[circle, thick, draw=brown, minimum size=40pt] (c1) at (-1.5,2.5){$\cO_5$};
        \foreach \i in {0,90,180,270} {
        \fill[black] (c1.\i) circle [radius=2pt]{};
        %\node at ($(c1.\i) + (0.5,0)$) {$P_1$}; % Place $P_1$ outside each circle
        \node at ($(c1.0) + (0.5,0)$) {$\bg_5$};
        \node at ($(c1.90) + (0,0.5)$) {$\bg_5+\bg_7$};
        \node at ($(c1.180) + (-1,0)$) {$\bg_5+\bg_6$};
        \node at ($(c1.270) + (0,-0.5)$) {$\bg_5+\bg_6+\bg_7$};
    }
    \end{tikzpicture}
    }
    \caption{Collection of row sets corresponding to $\Alow\cdot x_1.$}\label{fig:row-combin}%the orbit of monomial $f=x_1$, 
    %$\operatorname{Orb}(x_1)$ in Example \ref{ex:orbit_x1}.}
    \vspace{-10pt}
\end{figure}

\end{example}

%%%%%%%%%%%%%%%%%%%%%%%%%%%%%%%%%%%%%%%%%%%%%%%%%%%%%%%%%%%%%%%%%%%%%%%%

\section{General structure of codewords}\label{sec:wt_struct_enum}
%In the previous section, we observed that the orbit of a maximum-degree monomial $f$ resulting from the group action $\Alow\cdot f$ contains minimum-weight codewords. Obviously, 
When adding two minimum-weight codewords, we obtain a new codeword with Hamming weight in the interval $[\wm,2\wm]$. { It is possible to characterize all codewords of a decreasing monomial code with weight less than $2\wm$ using addition of multiple minimum-weight codewords \cite{dragoi2025polar}.} Therefore, one way of characterizing low-weight codewords is by looking at the sum of two orbits; $\Alow\cdot f+\Alow\cdot g$ where $f,g$ are maximum-degree monomials (see Fig.~\ref{fig:example1_intro}). In the literature, these are known as Type-I codewords \cite{kasami1970weight}. To build such codewords, we rely on the following. 
\begin{itemize}
    \item The Hamming weight of the sum of two codewords $\bc,\bc'$ is (inclusion-exclusion principle):
    \begin{equation}
        \w(\bc+\bc')=\w(\bc)+\w(\bc')-2\w(\bc\odot\bc'),
    \end{equation}
    where $\bc\odot\bc'$ is the component-wise product of $\bc$ with $\bc'.$ For example, the Hamming weight of $(0,1,1)+(1,1,0)$ equals 2 which equivalently is $\w(0,1,1)\!+\!\w(1,1,0)\!-\!\w(0,1,0)=2\!+\!2\!-\!2\!\cdot\! 1.$%Notice that $\bc*\bc'$ is a vector with support $\sup(\bc*\bc')=\sup(\bc)\cap\sup(\bc').$ 
    \item The component-wise product of two evaluation vectors $\ev(P),\ev(Q)$, where $P,Q$ are polynomials, %\in\ft[x_0,\dots,x_{m-1}]/\left\langle x_i^2-x_i\right\rangle$ 
    is the evaluation of their product
    \begin{equation}
        \ev(P)\odot\ev(Q)=\ev(PQ).
    \end{equation}
\end{itemize}

Applying these properties to $P\in\Alow\cdot f,Q=\Alow\cdot g$, $f,g\in\I_r$, we deduce 
\begin{equation}\label{eq:weight-sum-two}
        \w\left(\ev(P)+\ev(Q)\right)=2\cdot 2^{m-r}-2\cdot \w(\ev(PQ)).
    \end{equation}
%\vlad{Should I say that this uses inclusion-exclusion principle?}
The key idea to get a specific weight is to carefully select $P,Q$ such that we control $\w(\ev(PQ)).$ %For simplicity, we drop $H$ in $\w(\cdot)$.  %Let us provide some details about this construction.

\subsection{Type-I codewords} 
This type was studied in \cite{dragoi2025polar,dragoi2025weight}. Suppose we have $f,g\!\in\!\I_r$ and $P\!\in\!\Alow\cdot~f,Q=\Alow\cdot g$ as provided in \eqref{eq:weight-sum-two}. Since both $P,Q$ are product of $r$ independent linear forms (recall that linear forms are obtained from \eqref{eq:linear_form}), this means \eqref{eq:weight-sum-two} becomes 
\begin{equation}\label{eq:weight-sum-two2}
\w(\ev(P)+\ev(Q))=2^{m+1-r}-2^{m+1-\deg(PQ)},
\end{equation}
except for the case $\deg(PQ)=0$, where the degree of polynomial {$PQ$} is the degree of its largest monomial. 
%
%To determine the weight of the codeword corresponding to the evaluation of a polynomial $P=P_f+P_g$, where $P_f\in\Alow\cdot f,P_g\in\Alow\cdot g$, we need to compute the product $P_fP_g.$ This product is essential since it gives the overlapping of the two vectors $\ev(P_f),\ev(P_g)$ forming the codeword $\ev(P_f)+\ev(P_g).$ More exactly, we have \[\ev(P_f)+\ev(P_g)=\ev(P_f)+\ev(P_g)-2\times \ev(P_fP_g).\] 
%Let us give an example (see Fig.~\ref{fig:example1_intro}).
\begin{example} Let  $f=x_0x_1x_2,g=x_1x_3x_4, f,g\in\I$ for the polar code $\CI:(64,32)$. The weight of $P+Q$ for  $P\in\Alow\cdot f$ and $Q\in \Alow\cdot g$ equals:
\begin{itemize}
    \item $P=f=x_0x_1x_2$ and $Q=g=x_1x_3x_4.$ They overlap on $PQ=x_0x_1x_2x_3x_4$ which gives $\w(\ev(PQ))=2^{6-5}.$ Hence, $\w\left(\ev(P)+\ev(Q)\right)=2^{6-3}+2^{6-3}-2\cdot 2^{6-5}=2^{6-3+1}-2^{6-5+1}=12.$   
    %\item $P=x_0x_1x_2$ and $Q=(x_2+x_0)x_5x_6.$ They overlap on $PQ=(x_0+1)x_2x_3x_4x_5x_6$ which gives $\w(\ev(PQ))=2^{7-6}.$ Hence, $\w(PQ)=2^{7-3}+2^{7-3}-2\times 2^{7-6}=2^{7-3+1}-2^{7-6+1}.$  \mohammad{we can remove this one.} 
    \item $P=x_0x_1x_2$ and $Q=(x_1+1)x_3x_4.$ They overlap on $PQ=(x_1+x_1)x_0x_2x_3x_4=0$, i.e., they do not overlap at all. Hence, we obtain $\w\left(\ev(P)+\ev(Q)\right)=2^{6-3}+2^{6-3}.$     
\end{itemize}
\end{example}

Any Type-I codeword is fully determined by $f,g$ and the relation between $P$ and $Q$, specifically by the number of linear forms in $PQ.$ Let $\deg(PQ)=r+\mu$ represent this number where $\mu>0$ is a parameter  satisfying $m\geq r+\mu$ and $r\geq \mu\geq 3$ \cite{kasami1970weight}. Thus, the product $PQ$ consists of $r$ linear forms from $P$ and extra $\mu$ linear forms from $Q$, implying that $P,Q$ share $r-\mu$ common linear forms. Equivalently, $f,g$ have at least $r-\mu$ variables in common, denoted by $h=\gcd(f,g);$ 
\[ 
  \deg(PQ) \!=\! \deg(P)\!+\!\deg(Q)\!-\!\deg(h)
           \!=\! 2r \!-\! (r\!-\!\mu)
           \!=\! r\!+\!\mu.
\]
Consequently, the weight of the corresponding codeword as a function $\mu$, i.e., \eqref{eq:weight-sum-two2}. becomes
% \begin{equation}\label{eq:w_mu}
%     \boxed{\w_{\mu}=2^{m-r+1}-2^{m-r-\mu+1}=(2-\frac{1}{2^{\mu-1}})\wm.}
% \end{equation}  
\begin{equation}\label{eq:w_mu}
    %\fbox{$
    \w_{\mu}=2^{m-r+1}-2^{m-r-\mu+1}=(2-\frac{1}{2^{\mu-1}})\wm.%$}
\end{equation}  
% \begin{mdframed}[backgroundcolor=lightgray, linecolor=black, linewidth=1pt, roundcorner=0pt]
% \begin{equation}\label{eq:w_mu}
%     \w_{\mu}=2^{m-r+1}-2^{m-r-\mu+1}=(2-\frac{1}{2^{\mu-1}})\wm.
% \end{equation}  
% \end{mdframed}

\begin{figure*}[t]
\centering
\begin{tikzpicture}[
    x=0.7cm,y=0.7cm,
    >=latex,
    font=\scriptsize,
    box/.style={minimum height=0.5cm,anchor=west},
    lab/.style={anchor=west}
]

% ============= A1 =============
\begin{scope}
  \node[font=\footnotesize] at (2.2,3.3) {\textbf{A1: only $h$ is common}};

  % f bar
  \node[lab] at (-0.4,2.5) {$f$};
  \draw[fill=blue!25,draw=black] (0,2.3) rectangle (2.5,2.0);   % h
  \draw[fill=green!25,draw=black] (2.5,2.3) rectangle (4.5,2.0);% f/h

  % g bar
  \node[lab] at (-0.4,1.6) {$g$};
  \draw[fill=blue!25,draw=black] (0,1.35) rectangle (2.5,1.05);   % h
  \draw[fill=orange!25,draw=black] (2.5,1.35) rectangle (4.5,1.05);% g/h

  % Labels
  \node at (1.25,2.55) {$h$};
  \node at (3.5,2.55) {$f/h$};
  \node at (3.5,1.6) {$g/h$};

  \node[align=center] at (2.5,0.4) {
    clean overlap: $\gcd(f,g)=h$\\
    no extra restrictions
  };
\end{scope}

% ============= A2 =============
\begin{scope}[xshift=6cm]
  \node[font=\footnotesize] at (2.4,3.3) {\textbf{A2: extra common $h^\ast$}};

  % f bar
  \node[lab] at (-0.4,2.5) {$f$};
  \draw[fill=blue!25,draw=black]  (0,2.3) rectangle (1.8,2.0);  % h
  \draw[fill=red!25,draw=black]   (1.8,2.3) rectangle (3.1,2.0);% h*
  \draw[fill=green!25,draw=black] (3.1,2.3) rectangle (4.7,2.0);% f/(hh*)

  % g bar
  \node[lab] at (-0.4,1.6) {$g$};
  \draw[fill=blue!25,draw=black]  (0,1.35) rectangle (1.8,1.05);  % h
  \draw[fill=red!25,draw=black]   (1.8,1.35) rectangle (3.1,1.05);% h*
  \draw[fill=orange!25,draw=black](3.1,1.35) rectangle (4.7,1.05);% g/(hh*)

  % complement bar (for \widecheck{fg})
  \node[lab] at (-0.65,0.7) {$\widecheck{fg}$};
  \draw[fill=gray!15,draw=black]  (0,0.4) rectangle (4.7,0.1);
  \draw[fill=red!40,draw=black]   (0.8,0.4) rectangle (1.8,0.1);
  \node[text=red!70!black] at (1.3,0.68) {$h_s^\ast$};

  % Labels
  \node at (0.9,2.55) {$h$};
  \node at (2.45,2.55) {$h^\ast$};
  \node at (3.9,2.55) {$f/(hh^\ast)$};
  \node at (3.9,1.6) {$g/(hh^\ast)$};

  \node[align=center] at (2.5,-0.4) {
    $\gcd(f,g)=h\,h^\ast$ %; we keep only\\
    %LTA actions where $h^\ast$ becomes\\
    %independent via $J_{fg}(\cdot)$
  };
\end{scope}

% ============= B1 =============
\begin{scope}[xshift=12cm]
  \node[font=\footnotesize] at (2.4,3.3) {\textbf{B1: one monomial from frozen set}};

  % f bar (split into h and h*)
  \node[lab] at (-0.4,2.5) {$f$};
  \draw[fill=blue!25,draw=black] (0,2.3) rectangle (1.8,2.0);   % h
  \draw[fill=red!25,draw=black]  (1.8,2.3) rectangle (3.1,2.0); % h*
  \draw[fill=green!25,draw=black](3.1,2.3) rectangle (4.7,2.0); % f/(hh*)

  % highlight part outside I
  \draw[thick,dashed,red] (1.6,2.4) rectangle (4.9,1.9);
  \node[align=center,red] at (3.1,1.4) {dangerous  indices of $f$ whose children $x_jf/x_i$\\lie in $\I^c$ and must cancel in $P+Q$};%{monomials in $\mathcal I^c$\\ must cancel in $P+P'$};

  % information-set projection bar
  \node[lab] at (-0.7,0.6) {$J_f^{\mathcal I}$};
  \draw[fill=gray!15,draw=black]  (0,0.4) rectangle (4.7,0.1);
  \draw[fill=red!40,draw=black]   (0.8,0.4) rectangle (3.2,0.1);
  \node[text=red!70!black] at (2.4,0.65) {$\I$-valid indices of $f/h$};

  \node[align=center] at (2.4,-0.65) {
    $f=g\notin\mathcal I$; choose\\
    two orbit elements so that\\
    all terms in $\mathcal I^c$ cancel
  };
\end{scope}

% Legend (optional)
% \begin{scope}[yshift=-2.1cm]
%   \draw[fill=blue!25,draw=black] (0,0) rectangle +(0.7,0.35);
%   \node[lab] at (0.9,0.5) {$h$ (common factor of degree $r-\mu$)};

%   \draw[fill=red!25,draw=black] (5.0,0) rectangle +(0.7,0.35);
%   \node[lab] at (10,0.5) {$h^\ast$ (extra common part)};

%   \draw[fill=green!25,draw=black] (10.0,0) rectangle +(0.7,0.35);
%   \node[lab] at (17,0.5) {private part of $f$};

%   \draw[fill=orange!25,draw=black] (14.0,0) rectangle +(0.7,0.35);
%   \node[lab] at (22.5,0.5) {private part of $g$};
% \end{scope}
\begin{scope}[yshift=-2.1cm, xshift=0.3cm]
  \def\legy{1}
  \def\offset{3}
  % h
  \draw[fill=blue!25,draw=black] (0,\legy) rectangle +(0.7,0.35);
  \node[lab,anchor=west] at (0.9,\legy+0.175)
    {$h$ (common factor of degree $r-\mu$)};
  % h*
  \draw[fill=red!25,draw=black] (4.5+\offset,\legy) rectangle +(0.7,0.35);
  \node[lab,anchor=west] at (5.4+\offset,\legy+0.175)
    {$h^\ast$ (extra common part)};
  % private f
  \draw[fill=green!25,draw=black] (9.0+\offset+1,\legy) rectangle +(0.7,0.35);
  \node[lab,anchor=west] at (9.9+\offset+1,\legy+0.175)
    {private part of $f$};
  % private g
  \draw[fill=orange!25,draw=black] (13.5+\offset+1,\legy) rectangle +(0.7,0.35);
  \node[lab,anchor=west] at (14.4+\offset+1,\legy+0.175)
    {private part of $g$};
\end{scope}

\end{tikzpicture}
% \caption{Visualizing the three Type-I sub-types. Here, a bar = ``support of one monomial" or a set; the segments inside the bar = ``subsets of variables (indices) in that monomial." 
% In A1, $f$ and $g$ share only $h$ and both tails are free. In A2, they share an extra part $h^\ast$ and we restrict the LTA action so that the images of $h^\ast$ are independent (via $J_{fg}(\cdot)$). In B1, we start from a single frozen monomial $f=g\notin\mathcal I$ and choose two elements in its orbit so that all monomials in $\mathcal I^c$ cancel while the restriction based on $J_f^{\mathcal I}(\cdot)$ guarantees the Type-I weight.}
\caption{Type-I sub-types. Bars denote monomial supports (sets); segments denote variable subsets (indices). A1: $f$ and $g$ share only $h$ with free tails. A2: They share an extra part $h^\ast$; LTA restricts $h^\ast$ images to be independent via $J_{fg}(\cdot)$. B1: From a frozen $f=g\notin\mathcal I$, choosing two orbit elements cancels $\mathcal I^c$ monomials, while restriction based on $J_f^{\mathcal I}(\cdot)$ ensures Type-I weight.}
\label{fig:typeI-subtypes}
\end{figure*}

Observe that $1.5\leq(2-1/2^{\mu-1})<2$ for $\mu\geq 2$. 
For example, for the polar code $\CI\!:\!(64,32)$, we have 
\begin{center}
\setlength{\tabcolsep}{2pt}
\begin{tabular}{ c|ccccc } 
 $\mu$ & 1 & 2 & 3 \\%& 4\\% & 5\\ 
 \hline
 $\w_{\mu}$ & $\wm$ & $1.5\wm$ & $1.75\wm$ \\%& $1.875\wm$\\% & $1.9375\wm$ \\ 
 \hline
 Weight & 8 & 12 & 14 \\%& 120\\% & 31 \\ 
 %\hline
 %Type & {II} & II & II &  II\\% & I, II \\ 
\end{tabular}
\end{center}

Thus, as per \cite{dragoi2025polar}, %we know that %
%A complete characterization of Type-I codewords in terms of orbits under $\Alow$ was given in \cite{dragoi2025polar}. The results states that 
any Type-I codeword of weight $\w_{\mu}$ results from the evaluation of a polynomial in \begin{equation}
    \Alow_h\cdot h\left(\Alow_f\!\cdot\! \frac{f}{h}\!+\!\Alow_g\!\cdot\! \frac{g}{h}\right).
\end{equation} 
{For a fixed $\mu$, the target weight $\w_\mu$ is enforced via the constraint $\w(\deg(PQ))=r+\mu$. This requirement is met through various configurations of $f, g$, and $h$, governed by the arrangement of the shared component $h$ and the residual variables of $f$ and $g$ (and, when $f \notin I$, by enforcing cancellations within the frozen set). Fig.~\ref{fig:typeI-subtypes} illustrates these configurations, which will be described explicitly in the following. %The following sections make these configurations explicit. %and incorporate them into the unified counting formula \eqref{eq:typeI-unified}.
}

{
%\paragraph*{Unified counting formula for Type-I codewords}
%All three Type-I subcases (A1, A2, B1) can be written in a single form.  
For a triple $(f,g,h)$, generating Type-I codewords of weight $\w_\mu$, we let
$r = \deg(f) = \deg(g)$ and $\mu = r - \deg(h)$, we use the unified counting formula in \eqref{eq:typeI-unified} 
%\onecolumn
\vspace{-10pt}
\begin{multline}  
  A_{\w_\mu}^{\mathrm{I}}(f,g,h)
  = 2^{\,r+\mu - \delta_{\mathrm{sym}}
        + |\lambda_h|
        + \big|\lambda_f\!\big(\tfrac{f}{h}\big)\big|
        + \lambda_g^{\mathrm{free}}}\\
        \times
      \prod_{j=1}^{\deg(h^*)}
      \Big( 2^{\big|J^{\mathrm{eff}}(i_j)\big|} - 2^{j-1} \Big),
  \label{eq:typeI-unified}\vspace{-10pt}
\end{multline}
%\twocolumn
where %\vlad{page 8, col 1, line 44-51: The quantities $\delta_{\mathrm{sym}}, \lambda_g^{\mathrm{free}}, and J^{\mathrm{eff}}$ are introduced in a very compressed way.}{\color{red}We can just say that it is due to the space limitation.}
\begin{itemize}
  \item $h^*$ is the “extra” common factor (possibly trivial),
        with $\ind(h^*) = \{i_1,\dots,i_{\deg(h^*)}\}$,
  \item $\delta_{\mathrm{sym}}\in\{0,1\}$ (symmetry indicator, equal to $1$ when $f=g$)
  \item $\lambda_g^{\mathrm{free}}\in\mathbb{N}$ (number of remaining free affine
variables coming from $g/h$),
\item $J^{\mathrm{eff}}(\cdot)$ (effective sets of admissible lower indices for
the variables of $h^{*}$) are chosen according to the subcases in
Table~\ref{tab:typeI-unified}.
\end{itemize}

\begin{table}[t]
  \centering
  \resizebox{0.49\textwidth}{!}{
  
  \begin{tabular}{c|c|c|c}
    \toprule
    case & $h^*$ & $\lambda_g^{\mathrm{free}}$ & $J^{\mathrm{eff}}$ and $\delta_{\mathrm{sym}}$ \\
    \midrule
    A1 &
    $1$ &
    $\big|\lambda_g\!\big(\tfrac{g}{h}\big)\big|$ &
    $\deg(h^*) = 0$,\quad
    $\delta_{\mathrm{sym}}=0$
    \\[0.8ex]
    A2 &
    $\gcd\!\big(\tfrac{f}{h},\tfrac{g}{h}\big)$ &
    $\big|\lambda_g\!\big(\tfrac{g}{h}\big)\big| - \big|\lambda_{fg}h^*\big|$ &
    $J^{\mathrm{eff}} = J_{fg}$,\quad
    $\delta_{\mathrm{sym}} = \mathbf{1}_{\{f=g\}}$
    \\[0.8ex]
    B1 &
    $\tfrac{f}{h}$ &
    $0$ &
    $J^{\mathrm{eff}} = J_f^{\mathcal{I}}$,\quad
    $\delta_{\mathrm{sym}} = 1$
    \\
    \bottomrule
  \end{tabular}
  }
  \caption{Parameters for the unified Type-I formula~\eqref{eq:typeI-unified}.}
  \label{tab:typeI-unified}\vspace{-15pt}
\end{table}

With these choices (illustrated partially in Fig.~\ref{fig:typeI-subtypes}), Equation~\eqref{eq:typeI-unified} produces  three formulae:

%\begin{itemize}
\noindent\textbf{Case A1} (exactly $r\!-\!\mu$ common variables, $h \!=\!\gcd(f,g)$).
Here, $f$ and $g$ share only the factor $h$ of degree $r-\mu$, and their
quotients $f/h$ and $g/h$ have disjoint supports. Thus, there is no extra
common factor beyond $h$, so we set $h^{*}=1$ (and the product term in
\eqref{eq:typeI-unified} disappears), $\delta_{\mathrm{sym}}=0$ (since
$f\neq g$ in this case), and all affine variables on $g/h$ are free,
i.e.\ $\lambda_g^{\mathrm{free}} = \big|\lambda_g(\tfrac{g}{h})\big|$.
Plugging these choices into \eqref{eq:typeI-unified} gives
\begin{equation}
  A_{\w_\mu}^{\mathrm{A1}}(f,g,h)
  = 2^{\,r+\mu
        + |\lambda_h|
        + \big|\lambda_f\!\big(\tfrac{f}{h}\big)\big|
        + \big|\lambda_g\!\big(\tfrac{g}{h}\big)\big|}.
\end{equation}
Here, the exponent simply counts the binary degrees of freedom contributed by
the $r+\mu$ linear forms and the affine choices on $h$, $f/h$, and $g/h$.

    \begin{mdframed}[backgroundcolor=gray!10,linewidth=0pt]
        Type-I codewords for $\CI:(64,32)$\\
        For $\mu=3=r$, the only valid pair is of sub-type A.1; $f\!=x_0x_2x_5$, $g=x_1x_3x_4$ with $h=1$. This code does not have other sub-types. Thus, we have $\w_\mu=1.75\wm=14$ and 
        \[
        A_{14}(f,g,h)=2^{3+3+0+4+5}=2^{15}=32768.  %2^{3+4}2^{3+5}=2^{15}=32 768.
        \]
    \end{mdframed}

\textbf{Case A2} (more than $r-\mu$ common variables,
        $\gcd(f,g) = hh^*$, $h^*\neq 1$). 
The pair $(f,g)$ is valid if there exists a monomial $h_s^{*}$ such that:
\begin{itemize}
  \item \emph{Shift condition:} $h_s^{*}$ is a shift of $h^{*}$
    ($h_s^{*}\prec_{\mathrm{sh}} h^{*}$), i.e.,
    \[
      h_{s}^{*}=x_{i_1-\epsilon_1}\cdots x_{i_\ell-\epsilon_\ell},
      \qquad
      h^{*}=x_{i_1}\cdots x_{i_\ell},
    \]
    for some nonnegative shifts $\epsilon_1,\dots,\epsilon_\ell$.
  \item \emph{Complement condition:} $h_s^{*}$ divides the complement of $fg$,
    i.e.\ $h_s^{*} \mid \widecheck{fg}$.
\end{itemize}
Intuitively, we must be able to “slide” the extra common factor $h^*$ down to a
shift $h_s^*$ that fits entirely inside the positions not already used by $f$
and $g$.

\begin{mdframed}[backgroundcolor=gray!10,linewidth=0pt]
  For $\CI:(64,32)$ there is no valid pair of sub-type~A.2. For example,
  for $f=x_1x_3x_4$ and $g=x_1x_2x_4$, we have
  $\widecheck{fg}=x_0x_5$ and $h=1$, $h^{*}=x_1x_4$. Hence, there is no
  monomial $h_s^{*}$ such that $h_s^{*}\prec_{\mathrm{sh}}h^{*}$ and
  $h_s^{*}\preceq_w\widecheck{fg}$. The same obstruction occurs for all pairs
  in $\I_3$, so no A.2-type pairs exist for this code.
\end{mdframed}

Taking $\delta_{\mathrm{sym}}=\mathbf{1}_{\{f=g\}}$ (to encode the
“divide-by-$2$” symmetry when $f=g$) and
$\lambda_g^{\mathrm{free}}
       = \big|\lambda_g(\tfrac{g}{h})\big|-\big|\lambda_{fg}h^*\big|$
(to keep only the degrees of freedom in $g/h$ not already fixed by the extra
common factor $h^*$) in~\eqref{eq:typeI-unified}, the number of weight-$w_\mu$ codewords contributed by
Case~A2 is
\begin{equation}
\begin{aligned}
  A_{\w_\mu}^{\mathrm{A2}}(f,g,h)
    &= 
      \frac{
        2^{r+\mu-\mathbf{1}_{\{f=g\}} +|\lambda_h|
          +\big|\lambda_f(\tfrac{f}{h})\big|
          +\big|\lambda_g(\tfrac{g}{h})\big|}
      }{
        2^{|\lambda_{fg}h^*|}
      } \\
    &\qquad\times
      \prod_{j=1}^{\deg(h^*)}
        \Big(2^{|J_{fg}(i_j)|} - 2^{j-1}\Big).
\end{aligned}\vspace{-10pt}
\end{equation}
Here, $J_{fg}(i_j)$ denotes the set of admissible lower indices $k<i_j$ that can be
used to shift $x_{i_j}$ while preserving the A.2 validity conditions (in
particular, keeping the shifted copy $h_s^{*}$ inside $\widecheck{fg}$ and
maintaining full rank; see~\cite{dragoi2025polar}).
% \[
%   J_{fg}(i_j)
%   := \{\, k<i_j : x_k h^*/x_{i_j} \preceq_w \widecheck{fg}
%                   \text{ and the associated columns can be chosen to satisfy
%                   the full-rank condition on }\ind(h^*) \,\},
% \]
Hence, the product     $\prod_{j=1}^{\deg(h^*)}\big(2^{|J_{fg}(i_j)|} - 2^{j-1}\big)$ counts the  admissible matrices $\mathbf{B}$ in $(\mathbf{B},\varepsilon)\in\Alow_f$
    that satisfy the rank condition on the variables of $h^*$.

\begin{example}
Let $m=7,r=3,\mu=3$ and $f=x_3x_5x_6, g=x_1x_3x_5.$ Since $\deg(h)=r-\mu=0$, we have $h=1$. Thus, $h^{*}=x_3x_5$ and $fg=x_1x_3x_5x_6.$ The partition $\lambda_f=(6-2,5-1,3-0)=(4,4,3)$, visualized by the following Young diagrams: 
    \begin{center}
    \resizebox{0.3\textwidth}{!}{
    \begin{tikzpicture}[inner sep=0in,outer sep=5in]
      \node(m) at (0.5,0) {
\begin{ytableau}
 *(blue! 40) *  &*(blue! 40) *  & *(blue! 40) *  & 0  \\
 *(blue! 40) *  &*(blue! 40) *  & *(blue! 40) * & *(blue! 40) *   \\
*(blue! 40) *  &*(blue! 40) *  &*(blue! 40) *   &*(blue! 40) *  \\
\end{ytableau}};
\node at (-0.45,1.25) {$x_0$};
\node at (0.2,1.25) {$x_1$};
\node at (0.8,1.25) {$x_2$};
\node at (1.4,1.25) {$x_4$};
\node at (-1.1,0.6) {$x_3$};
\node at (-1.1,-0.05) {$x_5$};
\node at (-1.1,-0.7) {$x_6$};

\node(n) at (4.5,0){
\begin{ytableau}
 *(red! 40) \circ  &*(blue! 40) *  &*(red! 40) \circ   & 0 \\
 *(red! 40) \circ  & *(blue! 40) * & *(red! 40) \circ  &*(red! 40) \circ\\
*(blue! 40) *  &*(blue! 40) *   &*(blue! 40) * &*(blue! 40) *   \\
\end{ytableau}};
\node at (3.55,1.25) {$x_0$};
\node at (4.15,1.25) {$x_1$};
\node at (4.75,1.25) {$x_2$};
\node at (5.35,1.25) {$x_4$};
\node at (2.9,0.6) {$x_3$};
\node at (2.9,-0.05) {$x_5$};
\node at (2.9,-0.7) {$x_6$};
\end{tikzpicture}
}
\end{center}
Here, blue boxes represent binary choices for free variables and red boxes represent binary choices for linearly independent vectors. We have $\left|\Alow_f\cdot f\right|=2^{3+11}=2^{14}$ ($14$ blue-star boxes in the first Young diagram) and $\left|\Alow_f^g\cdot f\right|=2^{3+11-5}(2^2-1)(2^3-2)=2^93^2.$ 
The components are:
\begin{itemize}
    \item Translations: $2^{\deg(\frac{f}{h})}=2^{\deg(x_3x_5x_6)}=2^3$
    \item Free variables on $\frac{f}{h}$: $2^6$ ($6$ blue-star boxes in the second Young diagram)%$2^{|\lambda_f\frac{f}{h}|-|\lambda_{fg}h^{*}|}=2^{|\lambda_{x_3x_5x_6}x_3x_5x_6|-|\lambda_{x_1x_3x_5x_6}x_3x_5|}=2^{11-5}=2^{6}$
    \item Linearly independent forms of $h^{*}$: %since $\ind(h^{*})=\{x_3,x_5\}$ we have $J_{fg}(3)=\{0,2\},J_{fg}(5)=\{0,2,4\}$ which makes $\prod\limits_{j=1}^{\deg(h^{*})}\left(2^{|J_{fg}({i_j})|}-2^{j-1}\right)=
    $(2^{2}-2^0)(2^{3}-2^1)$ ($2$ red-circle boxes for $x_3$ and $3$ red-circle boxes for $x_5$ in the second Young diagram).
\end{itemize}
%\mohammad{Improve the clarity and coherence.}
%In total we have  
\end{example}

\noindent\textbf{Case B1} ($f=g\notin\mathcal{I}$ and frozen-set cancellations). 
In this case, we add two orbit elements of $f$,
\[
  P+Q = (B,\varepsilon)\cdot f + (B^{*},\varepsilon^{*})\cdot f,
\]
where $f\notin\mathcal I$. The monomial $f$ itself cancels in $P+Q$, so it is still possible that $\ev(P+Q)\in\C(\mathcal I)$. For this to happen, every degree-$r$ monomial appearing in $P+Q$ must belong to the information set $\mathcal I$. Equivalently, any degree-$r$ monomial in the orbit $\Alow\cdot f$ that lies in $\mathcal I^c$ must cancel in the sum
$(B,\varepsilon)\cdot f + (B^{*},\varepsilon^{*})\cdot f$.

This leads to the following conditions:
\begin{itemize}
  \item \textit{Existence of maximum-degree monomials:}\\
  For each $i\in\ind(f)$ there exists at least one $j<i$ with
  $j\notin\ind(f)$ such that
  $
    x_j f / x_i \in \mathcal I_r.
  $
   These indices form the sets $J_f^{\mathcal I}(i)$ and correspond to
  degree-$r$ children of $f$ that remain in $\mathcal I$.

  \item \textit{Cancellation of monomials in $\mathcal I^c$ (frozen set):}\\
  For any pair $(i,j)$ with $x_j f / x_i \notin \mathcal I$, we enforce
  $
    b_{i,j} = b_{i,j}^{*}
  $
   so that the corresponding monomial appears with the same coefficient
  in both orbit elements and hence cancels in $P$.
\end{itemize}
Under these constraints, all surviving degree-$r$ monomials of $P+Q$ lie in
$\mathcal I$, and thus $\ev(P+Q)\in\C(\mathcal I)$. %Note that the weight conditions on $f$ are identical to those in Case~A2. 

For counting, take $\lambda_g^{\mathrm{free}}=0$ (since $f=g$ gives no additional degrees of freedom from $g$), $h^{*}=f/h$ (so that only the high-degree indices of the quotient are active), and $J^{\mathrm{eff}} = J_f^{\mathcal I}$ (restricting to the $\mathcal I$-valid children of $f$), 
\[
  J_f^{\mathcal I}(i)
  := \{\, j<i : j\notin\ind(f),\ x_j f / x_i \in \mathcal I_r \,\},
\]
in~\eqref{eq:typeI-unified}, so the number of weight-$\w_\mu$ codewords contributed by Case~B1 is
%For counting, take $\lambda_g^{\mathrm{free}}=0$ (since $f=g$ contributes no independent degrees of freedom beyond those of $f$), $h^{*}=f/h$ (so the active high-degree positions in the product term are exactly the indices of the quotient on which the restricted orbit $\Alow_f^{f,\I}\!\cdot \tfrac{f}{h}$ acts), and $J^{\mathrm{eff}} = J_f^{\mathcal I}$ (encoding that, after the frozen-set cancellations $b_{i,j}=b_{i,j}^*$, only $\mathcal I$-valid descendants $x_j f/x_i$ remain free) in~\eqref{eq:typeI-unified}, so the number of weight-$w_\mu$ codewords contributed by Case~B1 is
  \begin{equation}
    A_{\w_\mu}^{\mathrm{B1}}(f,h)
      = 2^{r+\mu-1 + |\lambda_h|
             + \big|\lambda_f\big(\tfrac{f}{h}\big)\big|}
        \prod_{j=1}^{\mathclap{\deg(h^{*})}}
          \Bigl(2^{\big|J_f^{\mathcal I}(i_j)\big|} - 2^{j-1}\Bigr).
  \end{equation}

}
\begin{mdframed}[backgroundcolor=gray!10,linewidth=0pt]
    For $\CI:(64,32)$, there is no $f$ in the frozen set falling in the sub-case B.1. Let $f=x_1x_3x_5\not\in\I$ and $h=1.$% and $J_f(1)=\{0\},J_f(3)=\{0,2\},J_f(5)=\{0,2,4\}.$
    \vspace{-10pt}
    \begin{center}
    \resizebox{!}{0.35\columnwidth}{
    \begin{tikzpicture}[inner sep=0in,outer sep=5in]
      \node(m) at (0.5,0) {
\begin{ytableau}
 *(blue! 40) *  &    &   \\
 *(blue! 40) *  &*(blue! 40) *   &    \\
*(blue! 40) *  &*(blue! 40) *     &*(blue! 40) *  \\
\end{ytableau}};
\node at (-0.05,1) {$x_0$};
\node at (0.45,1) {$x_2$};
\node at (0.95,1) {$x_4$};
%\node at (1.25,1) {$x_4$};
%
\node at (-0.5-0.2,0.5) {$x_1$};
\node at (-0.5-0.2,0) {$x_3$};
\node at (-0.5-0.2,-0.5) {$x_5$};
\node at (0.45, 1.5) {$\lambda_f$};

%\node at (0.45,-1.25) {$x_2x_3x_5\not\in\I$};
\node(n) at (4.5,0){
\begin{ytableau}
 $X$  &&  \\
 *(red! 40) \circ  & $X$   &\\
*(red! 40) \circ  &*(red! 40) \circ   &$X$   \\
\end{ytableau}};
\node at (3.95,1) {$x_0$};
\node at (4.45,1) {$x_2$};
\node at (4.95,1) {$x_4$};
\node at (3.5-0.2,0.5) {$x_1$};
\node at (3.5-0.2,0) {$x_3$};
\node at (3.5-0.2,-0.5) {$x_5$};
\node at (4.45,1.5){$\I_3$};

\node at (4.45,-1.25){$x_2x_3x_5\not\in\I$};
\node at (4.45,-1.75){$x_1x_4x_5\not\in\I$};

\node(n) at (8.5,0){
\begin{ytableau}
 *(red! 40) \circ  &$X$& 0 \\
 *(red! 40) \circ  & *(red! 40) \circ   &$X$\\
*(red! 40) \circ  &*(red! 40) \circ   & *(red! 40) \circ   \\
\end{ytableau}};
\node at (3.95+4,1) {$x_0$};
\node at (4.45+4,1) {$x_1$};
\node at (4.95+4,1) {$x_3$};
\node at (3.5+4-0.2,0.5) {$x_2$};
\node at (3.5+4-0.2,0) {$x_4$};
\node at (3.5+4-0.2,-0.5) {$x_5$};
\node at (4.45+4,1.5){$\J_3$};
\node at (4.45+4,-1.25){$x_1x_4x_5\not\in\I$};
\node at (4.45+4,-1.75){$x_2x_3x_5\not\in\I$};
%\node at (4.45+4,-2.25){$x_2x_3x_4\not\in\I$};
\end{tikzpicture}
}.
\end{center}
The $X$-sign in the Young tableau denotes boxes that are fixed since the corresponding monomials are outside the set $\I.$ %Red boxes denote free variables that are taken such that the rank condition is satisfied. 

Since $x_0x_3x_5\not \in\I$ this implies that $J_f^{\I}(1)=\emptyset$ and thus $x_1x_3x_5$ is not a valid monomial.

For a higher-rate polar code (rate $0.6$) see the third diagram with information set $\J$, where $x_0x_4x_5,x_1x_2x_5,x_2x_3x_4\in\J_3$ suppose we consider a monomial $f=x_2x_4x_5\not\in\J.$ Since   $x_1x_4x_5,x_2x_3x_5\not\in\I$ we have $J_f^{\J}(2)=\{0\},J_f^{\J}(4)=\{0,1\},J_f^{\J}(5)=\{0,1,3\}$ and thus %we can apply the formula 
    \begin{itemize}
        \item $|\Alow_f\cdot f|=2^{3+1+2+3}=2^{13}$
        \item $|\Alow_f^{f,\J}\cdot f|=2^{3}(2^1-1)(2^2-2)(2^3-2^2)=2^{6}$
    \end{itemize}
\end{mdframed}

%%%%%%%%%%%%%%%%%%%%%%%%%%%%%%%%%%%%%%%%%%%%%%%%%%%%%%%

\subsection{Type-II codewords}
Type-II codewords, studied in \cite{rowshan2024weight,rowshan2025weight,vlad1.5d}, can be represented as a sum of $\mu$ polynomials, each corresponding to a minimum-weight codeword in the orbit $\Alow_f\cdot f$. %, in the following form 
% \begin{equation}
%     P=y_1\dots y_{r-2}(y_{r-1}y_{r}+\dots+y_{r+2\mu-3}y_{r+2\mu-2}).
% \end{equation}
% \mohammad{Maybe, we should not provide this equation as we have the LTA-based equation which looks simpler. If we remove it, we need to modify this paragraph a bit.} Recall that $y_i$ are linear independent forms obtained from \eqref{eq:linear_form}. %Such codewords $P$ are the result of LTA group actions on $\mu$ monomials $f_i\in\Mon$ of maximum-degree ($\deg(f_i)=r$) with $r-2$ shared variables for $1\leq \mu \leq \frac{m-r+2}{2}$ as
%satisfying the condition $m\geq r+\mu,r\geq \mu\geq 3$.  
%where each minimum-weight codeword belongs to an orbit $\Alow_f\cdot f$. %It was demonstrated in \cite{rowshan2024weight} that 
%These codewords %, as shown in \cite{rowshan2024weight}, 
%arises from evaluating a polynomial in
They correspond to polynomials 
% is generated by the action of an LTA group on $\mu$ monomials $f_i \in \I_r$ with $r-2$ shared variables, where $1 \leq \mu \leq \lfloor \frac{m-r+2}{2} \rfloor$,
\begin{equation}\label{eq:typeII-lta}
     \Alow_h\cdot h\left(\sum_{i=1}^{\mu}\Alow_{f_i} \cdot \frac{f_i}{h}\right),
\end{equation}
     where $\forall i, f_i\in \I_r$, $h=\gcd(f_i,f_j)$, for all $i,j\in[1,\mu]$ ($i\neq j$) and $\deg(h)=r-2$. The parameter $\mu$ satisfies $1 \leq \mu \leq \lfloor \frac{m-r+2}{2} \rfloor$. Note that $\mu=1$ gives minimum-weight codeword. %, which was studied in Section \ref{ssec:orbit_wm}. 
     Since $\deg(h)=r-2$, all the terms $f_i/h$ in the sum are degree-2 monomials ($\deg(f_i/h)=2$), making the study of their additive behavior a key aspect of understanding Type-II codewords. %, but before we detail this aspect let us continue looking at the Hamming weight. 
% \begin{equation}
%     \w_{\mu}(\ev(P))=(2-1/2^{\mu-1})\wm. 
% \end{equation}
% Observe that $1.5\leq(2-1/2^{\mu-1})<2$ for $\mu\geq 2$.
% For example, for the code $\R(2,8)\!:\!(256,37)$, we have 
% \begin{center}
% \setlength{\tabcolsep}{2pt}
% \begin{tabular}{ c|ccccc } 
%  $\mu$ & {1} & 2 & 3 & 4\\% & 5\\ 
%  \hline
%  $\w_{\mu}$ & {$\wm$} & $1.5\wm$ & $1.75\wm$ & $1.875\wm$\\% & $1.9375\wm$ \\ 
%  \hline
%  Weight & {64} & 96 & 112 & 120\\% & 31 \\ 
%  %\hline
%  %Type & {II} & II & II &  II\\% & I, II \\ 
% \end{tabular}
% \end{center}
%\vlad{Is the next figure useful? \mohammad{I think so.}}
The following figure demonstrates the structure of Type-II codewords {for a given $\mu$ value}.

\resizebox{0.9\columnwidth}{!}{
\begin{tikzpicture}
  % Common factor orbit (left, tight to the origin)
  \node[circle, thick, draw=brown, minimum size=46pt]
       (h1) at (0,2.5){\footnotesize$\Alow_h\cdot h$};

  % Multiplication sign, very close to the h-orbit
  \node at (1.7,2.5) {$\times$};

  % Sum of μ orbits of degree-2 quotients (label below)
  \node at (7.5,-0.2) {$\displaystyle \sum_{i=1}^{\mu}\Alow_{f_i}\cdot \frac{f_i}{h}$};

  % Rectangle covering all f_i-orbits, pulled close to the ×
  \draw (2.3,0.5) rectangle (12.7,4.5);

  % First orbit
  \node[circle, thick, draw=brown, minimum size=46pt, inner sep=1pt]
       (c11) at (3.7,2.5){\scriptsize$\Alow_{f_1}\cdot \frac{f_1}{h}$};

  % Second orbit
  \node[circle, thick, draw=brown, minimum size=46pt, inner sep=1pt]
       (c21) at (6.2,2.5){\scriptsize$\Alow_{f_2}\cdot \frac{f_2}{h}$};

  % Dots for intermediate orbits
  \node at (8.4,2.5) {$\cdots$};

  % Last orbit (generic f_μ)
  \node[circle, thick, draw=brown, minimum size=46pt, inner sep=1pt]
       (c31) at (10.8,2.5){\scriptsize$\Alow_{f_\mu}\cdot \frac{f_\mu}{h}$};

  % Sample points on the orbits
  \foreach \i in {20,60,100,140,180,220,260,300,340} {
    \fill[black] (c11.\i) circle [radius=2pt]{};
    \fill[black] (c21.\i) circle [radius=2pt]{};
    \fill[black] (c31.\i) circle [radius=2pt]{};
  }

\end{tikzpicture}
}

When adding two degree-2 polynomials, we may observe potential collisions, i.e., the sum of distinct polynomial pairs might be identical. This indicates that Type-II codewords can admit collisions. Specifically, for $f_1 = x_{i_1}x_{i_2}$ and $f_2 = x_{j_1}x_{j_2}$, there are three distinct cases to consider:
\begin{itemize}
    \item When $i_2 > i_1 > j_2 > j_1$: For example, $x_3x_2 + x_1x_0$ exhibits no collision (denoted $\alpha_{f_1, f_2} = 0$).
    \item When $i_2 > j_2 > i_1 > j_1$: For example, $x_4x_1 + x_2x_0 = (x_4 + x_0)x_1 + (x_2 + x_1)x_0$. In this case, each pair is counted twice (denoted $\alpha_{f_1, f_2} = 1$).
    \item When $i_2 > j_2 > j_1 > i_1$: Each pair is counted four times (denoted $\alpha_{f_1, f_2} = 2$). %\sout{See Fig.~\ref{fig:exemple-pack-collisions} for the case of $x_4x_2 + x_5x_0$ (denoted $\alpha_{f_1, f_2} = 2$).}
\end{itemize}

We need to discount identical polynomials resulting from collisions. %Dividing by $2^{\sum\limits_{(f_i,f_j)}\alpha_{\frac{f_i}{h},\frac{f_j}{h}}}$ is the standard way to correct for multiplicative overcounting in a binary choice space. 
Utilizing the collision exponent $\alpha_{f_1, f_2}$ defined above, and recalling $\lambda_{f}(g)$ from \ref{ssec:orbit_wm}, the formula for Type-II  is 
%\vlad{Equation (24) introduces $\alpha_{f_i,f_j} and \lambda_{f_i}(f_i/h)$ very abruptly. Even if these were defined earlier, the page would benefit from a brief reminder of what they mean.}
\vspace{-10pt}
\begin{equation}\label{eq:}
    A_{\w_{\mu}}^{\mathrm{II}}(f_i,i\in[1,\mu],h)\!=\!\dfrac{2^{r-2+2\mu+ |\lambda_h|+\sum\limits_{i=1}^{\mu}|\lambda_{f_i}(\frac{f_i}{h})|}}{2^{\sum\limits_{(f_i,f_j)}\alpha_{\frac{f_i}{h},\frac{f_j}{h}}}}.
\end{equation}
%The weight of Type-II codewords is computed as a function of $\mu$ using \eqref{eq:w_mu}, similarly to Type-I codewords. %$\w_{\mu}(\ev(P))=(2-1/2^{\mu-1})\wm$. 
%\sout{Table \ref{tab:formula-all} summarizes all formulae.}
%For any $\mu$, the corresponding formula is provided in Table~\ref{tab:formula-all}.
%W8 = 920, W12 = 25472, W14 = 32768.
\begin{center}
%\newlength{\mylength}
%\begin{figure}
% \setlength{\fboxsep}{5pt}
% \setlength{\mylength}{\linewidth}
% \addtolength{\mylength}{-3\fboxsep}
% \addtolength{\mylength}{-3\fboxrule}
% \shadowbox{
%     \parbox{\mylength}{
%     \setlength{\abovedisplayskip}{0pt}
%     \setlength{\belowdisplayskip}{0pt}
\begin{mdframed}[backgroundcolor=gray!10,linewidth=0pt]
    Type-II codewords for $\CI:(64,32)$\\
    Due to the constraint $\mu\!=\!\lfloor\frac{m-r+1}{2}\rfloor\!=\!2$, only monomial pairs are possible - We do not have triplets or more monomials for this code. Hence, according to \eqref{eq:w_mu}, the only weight we can have is
    %The computation of pairs $f_1/h,f_2/h$ for $h\!=\!x_0$ is given below. 
    $\w_{\mu}\!=\!1.5\wm\!=\!12$.\\ %\vspace{2pt}
       % \resizebox{0.4\texwidth}{!}{}
       \footnotesize
       \setlength{\tabcolsep}{4pt} % default is 6pt
 \begin{tabular}{c|c|c|c|c}
        \toprule
        $h$&$\frac{f_1}{h},\frac{f_2}{h}$ &   $\lambda_{f_1}(\frac{f_1}{h}),\lambda_{f_2}(\frac{f_2}{h})$ & $\alpha_{f_1,f_2}$ & $A_{12}(f_1,f_2,h)$\\
\midrule
         &$x_2x_5,x_3x_4$ & $(3,1),(2,2)$& 2&$2^{5+4+4-2}$\\
         &$x_2x_5,x_1x_4$ & $(3,1),(2,0)$& 1&$2^{5+4+2-1}$\\
         &$x_2x_5,x_1x_3$ & $(3,1),(1,0)$& 1&$2^{5+4+1-1}$\\
         &$x_2x_4,x_1x_5$ &$(2,1),(3,0)$& 2&$2^{5+3+3-2}$\\
         $x_0$&$x_2x_4,x_1x_3$ &$(2,1),(1,0)$& 1&$2^{5+3+1-1}$\\
         &$x_2x_3,x_1x_5$ &$(1,1),(3,0)$& 2&$2^{5+2+3-2}$\\
         &$x_2x_3,x_1x_4$ &$(1,1),(2,0)$& 2&$2^{5+2+2-2}$\\
         &$x_3x_4,x_1x_5$ &$(2,2),(3,0)$& 2&$2^{5+4+3-2}$\\
         &$x_3x_4,x_1x_2$ &$(2,2),(0,0)$& 0&$2^{5+4+0-0}$\\
         \midrule
         \multicolumn{5}{c}{\textbf{Total }$A_{12}\left(f_1, f_2, h=x_0\right)\!=\!6272.$}\\
         %\multicolumn{5}{c}{$\sum_{\substack{f_1, f_2 \in \I_3 \\ h\left|f_1, h\right| f_2}} A_{12}\left(f_1, f_2, h=x_0\right)\!=\!6272.$}\\
%           \bottomrule
%         %  \multicolumn{3}{c}{Total $A_{8}(\I)=920.$}\\
%         %  \bottomrule
%     \end{tabular}
% % }
% % }
% }
% \[
% \sum_{\substack{f_1, f_2 \in \I_3 \\ h\left|f_1, h\right| f_2}} A_{12}\left(f_1, f_2, h=x_0\right)\!=\!6272.
% \] 

%        {\footnotesize
%        \setlength{\tabcolsep}{4pt} % default is 6pt
%  \begin{tabular}{c|c|c|c|c}
%         \toprule
%         $h$&$\frac{f_1}{h},\frac{f_2}{h}$ &   $\lambda_{f_1}(\frac{f_1}{h}),\lambda_{f_2}(\frac{f_2}{h})$ & $\alpha_{f_1,f_2}$ & $A_{12}(f_1,f_2,h)$\\
%
\midrule
         \multirow{6}{*}{$x_1$}&$x_0x_5,x_3x_4$ & $(3,0),(2,2)$& 2&$2^{5+1+3+4-2}$\\
         &$x_0x_5,x_2x_4$ & $(3,0),(2,1)$& 2&$2^{5+1+3+3-2}$\\
         &$x_0x_5,x_2x_3$ & $(3,0),(1,1)$& 2&$2^{5+1+3+2-2}$\\
         &$x_0x_4,x_2x_3$ &$(2,0),(1,1)$& 2&$2^{5+1+2+2-2}$\\
         &$x_0x_3,x_2x_4$ &$(1,0),(2,1)$& 1&$2^{5+1+1+3-1}$\\
         &$x_0x_2,x_3x_4$ &$(0,0),(2,2)$& 0&$2^{5+1+4}$\\
         \midrule
          \multicolumn{5}{c}{\textbf{Total }$A_{12}\left(f_1, f_2, h=x_1\right)\!=\!5376.$}\\
          \midrule
            \multirow{4}{*}{$x_2$}&$x_0x_5,x_1x_4$ & $(3,0),(2,1)$& 2&$2^{5+2+3+3-2}$\\
         &$x_0x_5,x_1x_3$ & $(3,0),(1,1)$& 2&$2^{5+2+3+2-2}$\\
         &$x_0x_4,x_1x_3$ & $(2,0),(1,1)$& 2&$2^{5+2+2+2-2}$\\
         &$x_0x_3,x_1x_4$ &$(1,0),(2,1)$& 1&$2^{5+2+1+3-1}$\\
         \midrule
          \multicolumn{5}{c}{\textbf{Total }$A_{12}\left(f_1, f_2, h=x_2\right)\!=\!4608.$}\\
          \midrule
            \multirow{2}{*}{$x_3$}&$x_0x_2,x_1x_4$ & $(1,0),(2,1)$& 1&$2^{5+3+1+3-1}$\\
         &$x_1x_2,x_0x_4$ & $(1,1),(2,0)$& 2&$2^{5+3+2+2-2}$\\
         \midrule
          \multicolumn{5}{c}{\textbf{Total }$A_{12}\left(f_1, f_2, h=x_3\right)\!=\!3072.$}\\
          \midrule
            \multirow{2}{*}{$x_4$}&$x_0x_2,x_1x_3$ & $(1,0),(2,1)$& 1&$2^{5+4+1+3-1}$\\
         &$x_1x_2,x_0x_3$ & $(1,1),(2,0)$& 2&$2^{5+4+2+2-2}$\\
         \midrule
          \multicolumn{5}{c}{\textbf{Total }$A_{12}\left(f_1, f_2, h=x_4\right)\!=\!6144.$}\\
          \midrule
      %    
       %   \bottomrule
        %  \multicolumn{3}{c}{Total $A_{8}(\I)=920.$}\\
        %  \bottomrule
    \end{tabular}
% }
% }

% \[
% \sum_{\substack{f_1, f_2 \in \I_3 \\ h_1\left|f_1, h\right| f_2}} A_{12}\left(f_1, f_2, h=x_1\right)\!=\!5376.
% \] 
% Other divisors, $h\!=\!x_1, x_2, x_3, x_4, x_5$, also contribute to Type-II codewords, which appear in the following monomials:  %(for instance, $x_5$ appears only in $x_0 x_2 x_5, x_0 x_1 x_5$). 

% - $x_1$ divides $x_0x_1x_5, x_0x_1x_4, x_0x_1x_3, x_0 x_1 x_2, x_1 x_2 x_3,$ $x_1 x_3 x_4, x_1 x_2 x_4$. 

% - $x_2$ divides $x_0 x_2 x_5, x_0 x_2 x_4, x_0 x_2 x_3, x_0 x_1 x_2, x_1 x_2 x_3,$ $x_1 x_2 x_4$. 

% - $x_3$ divides $x_0 x_2 x_3, x_0 x_1 x_3, x_1 x_2 x_3, x_1 x_3 x_4, x_0 x_3 x_4$. 

% - $x_4$ divides $x_0 x_2 x_4, x_0 x_1 x_4, x_1 x_3 x_4, x_1 x_2 x_4, x_0 x_3 x_4$. 

% - $x_5$ divides $x_0 x_2 x_5, x_0 x_1 x_5$. 

%\medskip
%\vspace{2pt}
% We need to compute $A_{12}(f_1,f_2,h)$ for all ($f_1,f_2$)-pairs sharing $h$ similar to the table above, and sum them up. The remaining number of codewords is 
\[
A_{12}=\sum_{h\in\{x_i,i\in[0,4]\}}\sum_{\substack{f_1, f_2 \in \I_3 \\ h\left|f_1, h\right| f_2}}A_{12}(f_1,f_2,h)\!=\!25472.
\] 
% which results in total of
% $$A_{12}\!=\!6272 \!+\! 19200 \!=\! 25472.$$
%We need to find all monomials in $\I_r$ satisfying the conditions in Table \ref{tab:formula-all} and calculate the cardinality of the corresponding orbits. The sum of all $A_{\w_{\mu}}$ of all types gives the following for the $\CI:(64,32)$ polar code:
Thus, the partial weight distribution of $\CI:(64,32)$ is: 
\[A_8 = 920\;,\; A_{12} = 25472 \;,\; A_{14} = 32768.\]
\end{mdframed}
    
\end{center}

\section{Enumeration of Shortened Codes}\label{sec:wt_short_code}
%ISIT paper
Let $\C(\mathcal{I})$ be a polar code with length $N=2^m$, and $\S \subseteq[N]$ be a set called a shortening pattern. The shortened code with a shortening pattern $\S$ is $\C_s(\mathcal{I}, \S)=\left\{\bc_{[N] \setminus \S} \mid \bc \in\C(\mathcal{I}), \bc_\S=0\right\}$. That is, a shorten code is obtained by setting certain bits (defined by a shortening pattern $\S$) to zero, effectively removing the codewords with $\bc_\S\not=0$ from the code and reducing the length to $N-|\S|$. 

Since the codewords in $\C(\I, \S)$ are those in $\C(\I)$ that are zero on the coordinates in $\S$, we take only the polynomials $P$ from the original set that evaluate to zero, $P(x_0,\ldots,x_{m-1})=0$ at all points $(u_0,\dots ,u_{m-1})$ in $\S$. %\vlad{It would be easier if points are denoted with $(u_0,\dots ,u_{m-1})$ since $x_i$ stands for variable}. 
If $\S$ contains a coordinate corresponding to monomial $g$, we set the variables in $g$ to 1 and others to 0. For example, if we shorten the last bit, the corresponding monomial and point are $g=1$ and $(0,0,\ldots,0)$. 
Now, considering multiple shortened coordinates/monomials, a system of equations emerges \cite{ye2024weight}. %The bit-reversal shortening has an equation system that is solvable, while for other shortening approaches, solving the equations system might be challenging. 

%Denote the bit-reversal sequence $q^{\prime}$ to be the bit reversal of $\boldsymbol{q}$, i.e., $\boldsymbol{q}_{D\left(a_m, \ldots, a_1\right)}^{\prime}=\boldsymbol{q}_{D\left(a_1, \ldots, a_m\right)}=D\left(a_1, \ldots, a_m\right)$. 

%The bit-reversal shortening pattern $\S$ is the last $|\S|$ bits in bit-reversal order, and the bit-reversal shortened polar code \cite{bioglio2017low} with length $N-|\S|$ and information set $\mathcal{I}$ is denoted as $\C\left(\mathcal{I}, \S\right)$. 
%Given the codeword polynomials $P$ in $\Orb(f)$, when the monomial $g$ is shortened, a codeword $P(x_0,\ldots,x_{m-1}=\sum_{g \in \I} a_g g$ survives if and only if the evaluation of $P$ at the point where the variables in $g$ are 1 and others are 0 , requiring $\vec{P}$ to evaluate to 0 at that point, and then removing that coordinate from the codeword.

%the coefficient $a_g=0$ for $g \in \S$. This imposes a constraint equation on the relevant coefficients $a_{j, s}$ or $b_j$. Now, considering multiple shortened monomials, a system of equations emerges. The bit-reversal shortening has the equation system which is indeed solvable. 

\begin{example}
%\vlad{Attention: $N=32$ is for $m=5$ the variable $x_5$ does not exist!}
    Assume $N=32, f=x_2 x_4$. Then, all codewords in the orbit of $f$ %, $\Orb(f)$, 
    are 
    $$
    \begin{aligned}
        P= & \left(x_4+a_{4,3} x_3+a_{4,1} x_1+a_{4,0} x_0+b_4\right) \\
        & \cdot\left(x_2+a_{2,1} x_1+a_{2,0} x_0+b_2\right)
    \end{aligned}
    $$
    The first shortened bit is the last bit, $u_{31}\!=\!c_{31}\!=\!0$, corresponding to the monomial $1$ and the point $(0,0,0,0,0)$. Evaluating $P$ results in:
$$
    \begin{aligned}
        P(0,0,0,0,0)= & \left(0+a_{4,3} \cdot 0+a_{4,1} \cdot 0+a_{4,0} \cdot 0+b_4\right) \\
        & \cdot\left(0+a_{2,1} \cdot 0+a_{2,0} \cdot 0+b_2\right)=b_4 b_2.
    \end{aligned}
$$
    Hence, for $P$ to survive, we need $b_4 b_2=0$, meaning $b_4=0, b_2=0$, or both. This removes all orbit members where $b_4=b_2=1$, which is a quarter of all four possible combinations for $b_4,b_2\in\{0,1\}$. Hence, assuming $A_{\min}$ for the mother code, the multiplicities of the shortened code will be $\frac{3}{4}A_{\min}$. 
    
    Next, the monomial $x_4$ corresponding to evaluating point $(0,0,0,0,1)$ is shortened, i.e., $u_{15}\!=\!c_{15}\!=\!0$ in bit-reversal order, that results in
$$
\begin{aligned}
    P(0,&0,0,0,1)=\left(1\!+\!a_{4,3} \cdot 0\!+\!a_{4,1} \cdot 0\!+\!a_{4,0} 
    \cdot 0\!+\!b_4\right) \\
    &\cdot \left(0+a_{2,1} \cdot 0+a_{2,0} \cdot 0+b_2\right)=\left(1+b_4\right) b_2.
\end{aligned}
$$
For $P$ to survive, we need $\left(1+b_4\right) b_2=0$ which is satisfied if $b_2=0$, or $1+b_4=0$, i.e., $b_4=1$. Since the case $b_4=b_2=1$ is already shortened, the newly removed polynomials are those with $b_4=1, b_2=0$. Of four combinations for $b_4,b_2\in\{0,1\}$, the removed polynomials are those with $b_4=0, b_2=1$, as they yield $(1+0)$. $1=1$. Hence, another quarter of min-weight codewords is removed, and the remaining after two bit shortening will be $\frac{1}{2}A_{w_{\min}}$.

    %which demands $b_5=0$, polynomials with $b_5=1$ are no longer codewords. Since the case $b_5=b_3=1$ is already shortened, the newly removed polynomials are those with $b_5=1, b_3=0$. 
    %Thus, $\lambda_f / 4$ more minimum-weight codewords are removed, and the remaining number of codewords with weight 8 in $\Orb(f)$ is $\lambda_f / 2$ after 2-bit bit-reversal shortening.
\end{example}

%%%%%%%%%%%%%%%%%%%%%%%%%%%%%%%%%%%%%%%%%%%%%%%%%%%%%%%%%%%%%%%%%%%%%%%%
\section{Recursive Weight Structure}\label{sec:recursive_enum}
%Vardy's paper
%{\color{red} Here we use $G_{2N}$ it should be in bold. Maybe simplify this first part where the bit reversal permutation is used.}
Given the recursive structure of the polar transform \cite[Proposition 3]{Arikan}, we can construct every codeword $\bc=\bu \cdot \mathbf{G}_{2N}$ by two codewords $\bc_1$ and $\bc_2$ of half-length. {This key idea underpins the recursive weight enumeration of polar codes within cosets, as first proposed in \cite{yao}.} %$N/2=2^{m-1}$ as follows. 
Note that $\bG_{N}$ is full-rank and any sum of $\boldsymbol{g}_i$ is bijective (one-to-one and onto). Denote $\boldsymbol{u}_{\text{even}}=(u_0,u_2,\dots,u_{2N-2})$, $\boldsymbol{u}_{\text{odd}}=(u_1,u_3,\dots,u_{2N-1})$, and $\bB_{2N}$ the bit-reversal matrix, we have %{\color{red} Do we really need this part with the bitreversal and the F matrix? We go directly to the last two equations.}
%codeword $\bc$ as %\vlad{$\boldsymbol{u}_{\text{even}}$ and $B_{2N}$ to be defined}
\begin{equation}\footnotesize
\begin{aligned}
\bc=\boldsymbol{u} \cdot \bG_{2N} & =\left(\boldsymbol{u} \cdot \bB_{2N}\right) \bG_2^{\otimes(m+1)} \\
& =\left(\boldsymbol{u}_{\text{even}} \cdot \bB_{N}, \boldsymbol{u}_{\text{odd}} \cdot \bB_{N}\right)\left[\begin{array}{cc}
\bG_2^{\otimes m} & 0 \\
\bG_2^{\otimes m} & \bG_2^{\otimes m}
\end{array}\right] \\
& =\left(\left(\boldsymbol{u}_{\text{even}} \oplus \boldsymbol{u}_{\text{odd}}\right) \cdot \bB_{N} \bG_2^{\otimes m}, \boldsymbol{u}_{\text{odd}} \cdot \bB_{N} \bG_2^{\otimes m}\right) \\
& =(\underbrace{\left(\boldsymbol{u}_{\text{even}} \oplus \boldsymbol{u}_{\text{odd}}\right) \cdot \bG_{N}}_{\bc_1}, \underbrace{\boldsymbol{u}_{\text{odd}} \cdot \bG_{N}}_{\bc_2}).
\end{aligned}
\end{equation}
Splitting $\bu$ into $\bu=\left(\bu_{2i},u_{2i+1},\bv\right)$ where $\bu_{2i}=u_0^{2i}$ consists of an odd number of elements, the inclusion of $u_{2i+1}$ ensuring the equal number of odd and even indices for recursion, and $\bv\in\{0,1\}^{2N-2i-2}$, %\vlad{The size of $\bm{u}$ should be $2N$} 
we can rewrite the two codewords $(\bc_1,\bc_2)$ as
$$
\bc_1=\left(\bu_{2 i, \text{even}} \oplus\left(\bu_{2 i, \text{odd}}, u_{2i+1}\right), \bv_{\text{even}} \oplus \bv_{\text{odd}}\right) \cdot \mathbf{G}_{N}
$$
%and
$$
\bc_2=\left(\bu_{2i, \text{odd}}, u_{2i+1}, \bv_{\text{odd}}\right) \cdot \mathbf{G}_{N},
$$
where $u_{2i+1}\in\{0,1\}$. 
%Note that the length of the vectors changes. In particular, while we have $\bv\in\{0,1\}^{2N-2i-1}$, after splitting , we get $\bv_{\text{odd}},\bv_{\text{even}}\oplus\bv_{\text{odd}}\in\{0,1\}^{N-i-1}$.

Now, if we define a coset as  $$C_N\left(\bu_i\right)=\left\{\left(\bu_i, \bv\right) \mathbf{G}_N \mid \bv \in\{0,1\}^{N-i-1}\right\},$$ where the first $i+1$ bits ($u_0,u_1,\dots,u_i$) are fixed, %and $C_N\left(\bu_i\right)\subset C_N\left(\bu_{i-1}\right)$ \cite{ellouze2024distance}, 
we can split every coset into two sub-cosets as
\begin{equation}\label{eq:CNu_i+1&}
\begin{aligned}
& C_{2N}\left(\boldsymbol{u}_{2 i}, u_{2 i+1}\right)= \\
& \quad\left\{\left(\boldsymbol{c}_1, \boldsymbol{c}_2\right) \mid \boldsymbol{c}_1 \in C_N\left(\boldsymbol{u}_{2 i, \text{even}} \oplus\left(\boldsymbol{u}_{2 i, \text{odd}}, u_{2 i+1}\right)\right)\right., \\
& \left.\quad \boldsymbol{c}_2 \in C_N\left(\boldsymbol{u}_{2 i, \text{odd}}, u_{2 i+1}\right)\right\},
\end{aligned}
\end{equation}
which shows %for each $u_{2i+1}\in\{0,1\}$, 
the coset $C_{2N}\left(\boldsymbol{u}_{2 i}, u_{2 i+1}\right)$ is the direct sum of two cosets $C_{N}\left(\cdot\right)$. {The notion of coset used here generalizes the definition in \cite{polyanskaya}.}

The weight enumerating function (WEF) of a code or coset is a polynomial that counts the number of codewords of each Hamming weight. For a coset $C_{2N}\left(\bu_i\right)$, the WEF is:
$$
A_{2N}\left(\bu_i\right)(X)=\sum_{\bc \in C_{2N}\left(\bu_i\right)} X^{\w(\bc)},
$$
where the sum is over all codewords in the coset. The coefficient of $X^k$ in $A_{2N}\left(\bu_i\right)(X)$ will be the number of codewords with Hamming weight $k$. 

%- WEF for Direct Sum: The statement claims that the WEF of the direct sum of two polar cosets equals the product of their individual WEFs. This is a property that arises 
For a codeword $\left(\bc_1, \bc_2\right) \in C_{2N}\left(\bu_{2 i}, u_{2 i+1}\right)$, the Hamming weight is: 
$$
\w\left(\left(\bc_1, \bc_2\right)\right)=\w\left(\bc_1\right)+\w\left(\bc_2\right).
$$
This additive property suggests that the weights of $\bc_1$ and $\bc_2$ contribute independently to the total weight, which indicates the WEF of the direct sum of two cosets equals the product of their two individual WEFs. 

The coset $C_{2N}\left(\bu_{2 i}\right)$ can be expressed as the union of two cosets by considering the bit $u_{2i+1}$:
$$
C_{2N}\left(\bu_{2 i}\right)=C_{2N}\left(\bu_{2 i}, 0\right) \cup C_{2N}\left(\bu_{2 i}, 1\right)
$$
where $C_{2N}\left(\bu_{2 i}, u_{2 i+1}\right)$ fixes the first $2i+1$ bits to $\bu_{2 i}$ and the $(2 i+1)$-th bit to $u_{2 i+1}$. These two cosets are disjoint because fixing $u_{2 i+1}=0$ or $u_{2 i+1}=1$ produces different sets of codewords.

The WEF of $C_{2N}\left(\boldsymbol{u}_{2 i}\right)$ is the sum of the WEFs of the two sub-cosets:
\begin{equation}\label{eq:enum_even}
\begin{aligned}
A_{2N}\left(\bu_{2 i}\right)(X) & =\sum_{\bc \in C_{2N}\left(\bu_{2 i}\right)} X^{\w(\bc)}\\ & =\sum_{u_{2 i+1} \in\{0,1\}} \sum_{\bc \in C_{2N}\left(\bu_{2 i}, u_{2 i+1}\right)} X^{\w(\bc)} \\
& =\sum_{u_{2 i+1} \in\{0,1\}} A_{2N}\left(\bu_{2 i}, u_{2 i+1}\right)(X).
\end{aligned}
\end{equation}

Each codeword $\left(\bc_1, \bc_2\right)$ has weight $\w\left(\bc_1\right)\!+\!\w\left(\bc_2\right)$; that is, $X^{\w\left(\bc\right)}=X^{\w\left(\bc_1\right)+\w\left(\bc_2\right)}$. %the weights combine additively in the exponent of $X$. 
Then, the WEF of each sub-coset in \eqref{eq:enum_even} for a fixed $u_{2 i+1}$ is \cite{yao}
% $$
% A_{2N}\left(\bu_{2 i}, u_{2 i+1}\right)(X)=\smashoperator{\sum_{\left(\bc_1, \bc_2\right) \in C_{2N}\left(\bu_{2 i}, u_{2 i+1}\right)}} X^{\w\left(\bc_1\right)+\w\left(\bc_2\right)}
% $$
\begin{equation}\label{eq:enum_even_sub}
\begin{aligned}
A_{2N}  \left(\bu_{2 i}, u_{2 i+1}\right)(X)%&=\\
%\sum_{\bc_1 \in C_N\left(\bu_{2i, \text{even}} \oplus\left(\bu_{2 i, \text{odd}}, u_{2 i+1}\right)\right)} \sum_{\bc_2 \in C_N\left(\bu_{2, \text{odd}}, u_{2 i+1}\right)} X^{\w\left(\bc_1\right)+\w\left(\bc_2\right)} \\
& = \smashoperator{\sum_{\bc_1 \in C_N\left(\bv\right)} } X^{\w\left(\bc_1\right)}   \cdot \smashoperator{\sum_{\bc_2 \in C_N\left(\bw\right)}} X^{\w\left(\bc_2\right)} \\
& =A_N\left(\bv\right)(X) \cdot A_N\left(\left(\bw\right)\right)(X)
\end{aligned}
\end{equation}
where the prefix vectors are $\bv=\bu_{2 i, \text{even}} \oplus\left(\bu_{2 i, \text{odd}}, u_{2 i+1}\right)$ and $\bw=\left(\bu_{2 i, \text{odd}}, u_{2 i+1}\right)$. 
Note that since $\bc_1$ and $\bc_2$ are chosen independently from their respective cosets $C_N\left(\bv\right)$ and $C_N\left(\bw\right)$, the WEF is the product of the individual WEFs.
% $$
% \begin{aligned}
% A_{2N}  \left(\bu_{2 i}, u_{2 i+1}\right)(X)%&=\\
% %\sum_{\bc_1 \in C_N\left(\bu_{2i, \text{even}} \oplus\left(\bu_{2 i, \text{odd}}, u_{2 i+1}\right)\right)} \sum_{\bc_2 \in C_N\left(\bu_{2, \text{odd}}, u_{2 i+1}\right)} X^{\w\left(\bc_1\right)+\w\left(\bc_2\right)} \\
% & = \smashoperator{\sum_{\bc_1 \in C_N\left(\bv\right)} } X^{\w\left(\bc_1\right)}   \cdot \smashoperator{\sum_{\bc_2 \in C_N\left(\bw\right)}} X^{\w\left(\bc_2\right)} \\
% & =A_N\left(\bv\right)(X) \cdot A_N\left(\left(\bw\right)\right)(X)
% \end{aligned}
% $$
%This product form arises because the codewords $\bc_1$ and $\bc_2$ are generated independently, and their weights combine additively in the exponent of $X$. 

% the coset $C_{2N}\left(\boldsymbol{u}_{2 i}, u_{i+1}\right)$ consists of all vectors $\left(\boldsymbol{c}_1, \boldsymbol{c}_2\right)$ can be expressed as the direct sum of two polar cosets
% \begin{equation}
%     \boldsymbol{c}_1 \in C_n\left(\boldsymbol{u}_{2 i, \text{even}} \oplus\left(\boldsymbol{u}_{2 i, \text{odd}}, u_{2 i+1}\right)\right),
% \end{equation}
% and
% \begin{equation}
%     \boldsymbol{c}_2 \in C_n\left(\boldsymbol{u}_{2 i, \text{odd}}, u_{2 i+1}\right).
% \end{equation}

Now, let us rewrite \eqref{eq:CNu_i+1&} as 
\begin{equation}\label{eq:CNu_i+1}
\begin{aligned}
& C_{2N}\left(\boldsymbol{u}_{2 i+1}\right)= \\
& \quad\left\{\left(\bc_1, \bc_2\right) \mid \bc_1 \in C_N\left(\boldsymbol{u}_{2 i+1, \text{even}} \oplus \boldsymbol{u}_{2 i+1, \text{odd}}\right),\right. \\
& \left.\quad \bc_2 \in C_N\left(\boldsymbol{u}_{2 i+1, \text{odd}}\right)\right\}. 
\end{aligned}
\end{equation}
Then, similarly, the WEF is the product of the individual WEFs as \cite{yao}
\begin{equation}\label{eq:enum_odd}
\begin{gathered}
A_{2N}\left(\bu_{2 i+1}\right)(X)=\sum_{\bc_1 \in C_N\left(\bv'\right)} X^{\w\left(\bc_1\right)} \cdot \sum_{\mathclap{\bc_2 \in C_N\left(\bw'\right)}} X^{\w\left(\bc_2\right)} \\
=A_N\left(\bv'\right)(X) \cdot A_N\left(\bw'\right)(X)
\end{gathered}
\end{equation}
where $\bv'=\bu_{2 i+1, \text{even}} \oplus \bu_{2 i+1, \text{odd}}$ and $\bw'=\bu_{2 i+1, \text{odd}}$. 

We can use the recursive relations \eqref{eq:enum_even} and \eqref{eq:enum_odd} for weight enumeration of a coset as follows: %, in particular, if we consider $u_{2i+1}$ as the information bit following the last frozen bit. 
In each recursive step, to compute $A_{2 N}\left(\bu_i\right)(X)$ for $\bu_i\in\{0,1\}^{i+1}, 0\leq i\leq N-1$ and $N=2^m$, 
\begin{itemize}
    \item if $i=2 i^{\prime}$ (even), $\mathbf{u}_{2 i^{\prime}}$ has $2 i^{\prime}+1$ fixed bits, with $i^{\prime}+1$ even-indexed bits and $i^{\prime}$ odd-indexed bits (as the bit $u_{2i'+1}\in\{0,1\}$ is a free bit). In this case, we use \eqref{eq:enum_even}.
    \item if $i=2 i^{\prime}+1$ (odd), $\mathbf{u}_{2 i^{\prime}+1}$ has $2 i^{\prime}+2$ fixed bits, with $i^{\prime}+1$ bits each in the even and odd parts. In this case, we use \eqref{eq:enum_odd}.
\end{itemize} 
%The bit $u_{2 i^{\prime}+1}$ appears in the odd part ($u_{\text{odd}}$), affecting both $c_1$ and $c_2$. 
Note that the difference in these two cases is the number of fixed bits. 
%The bit $u_{2 i^{\prime}+1}$ is included in $\mathbf{u}_{2 i^{\prime}+1, \text {odd}}$ and affects the XOR in $\mathbf{u}_{2 i^{\prime}+1, \text{even}} \oplus \mathbf{u}_{2 i^{\prime}+1, \text{odd}}$. 
Each recursive call reduces $N$ to $N / 2$, and the index $i^{\prime}$ is approximately halved (since $i^{\prime} = \lceil (i-1) / 2\rceil$). The process continues until $N=1$, which is the base case in the recursion where $i=0, u_0 \in\{0,1\}$, and we have  
\[
A_1(u_0)(X) = \begin{cases}
    1 & \text{if } u_0 = 0, \\
    X & \text{if } u_0 = 1.
\end{cases}
\]

\begin{example}\label{ex:recursive_enum}
%1     0     8     0    14     0     8     0     1
   Consider the polar code $(8,6)$ with $\A = [2,7]$. Knowing $i=\max(\A^c)=1$, we compute $A_8(\mathbf{u}_1)(X)$ recursively, beginning from $N = 8$ and decomposing through $N = 4$, $N = 2$, and $N = 1$, as shown in Fig.~\ref{fig:coset_decomposition}, then reconstructing the solution. Since $i = 1$ is odd ($i = 2i' + 1$ with $i' = 0$), we apply \eqref{eq:enum_odd}. Note that $u_0^1$ contains no free bits; thus, a single coset determines the entire weight distribution. 
    \[\footnotesize
    \begin{aligned}
        A_8\left(\bu_1\right)(X)&=  A_4\left(u_0\oplus u_1\right)(X)\cdot A_4\left(u_1\right)(X) \\
        A_8\left([0,0]\right)(X)&= A_4\left(0\right)(X)\cdot A_4\left(0\right)(X)=\left(A_4\left(0\right)(X)\right)^2.
    \end{aligned}
    \]
    Since the new $i=0$ for $A_4$ is even $\left(i=2 i^{\prime}, i^{\prime}=0\right)$, we use $\eqref{eq:enum_even}$:
    \[\footnotesize
    \begin{aligned}
        A_4\left(u^{\prime}_0\right)(X)&=  A_2\left(u^{\prime}_0\oplus 0\right)(X)\cdot A_2\left(0\right)(X) \\&+ A_2\left(u^{\prime}_0\oplus 1\right)(X)\cdot A_2\left(1\right)(X)\\ 
        A_4\left(0\right)(X)&
        =A_2\left(0\right)(X)\cdot A_2\left(0\right)(X)+ A_2\left(1\right)(X)\cdot A_2\left(1\right)(X).
    \end{aligned}
    \]
    Now, compute the WEFs for $N=2$. Here, $i=0$, which is even $\left(i=2 i^{\prime}, i^{\prime}=0\right)$, so we use $\eqref{eq:enum_even}$:
    \[\footnotesize
    \begin{aligned}
        A_2\left(u^{\prime\prime}_{0}\right)(X)&=A_1\left(u^{\prime\prime}_0\oplus 0\right)(X) \cdot A_1\left(0\right)(X)\\&+A_1\left(u^{\prime\prime}_0\oplus 1\right)(X) \cdot A_1\left(1\right)(X),\\
        %A_2\left(0\right)(X)&=A_1\left(0\oplus 0\right)(X) \cdot A_1\left(0\right)(X)\\&+A_1\left(0\oplus 1\right)(X) \cdot A_1\left(1\right)(X),\\
        A_2\left(0\right)(X)&=A_1\left(0\right)(X) \cdot A_1\left(0\right)(X)+A_1\left(1\right)(X) \cdot A_1\left(1\right)(X)\\&=1\cdot1+X\cdot X=1+X^2,\\
        A_2\left(1\right)(X)&=A_1\left(1\right)(X) \cdot A_1\left(1\right)(X)+A_1\left(0\right)(X) \cdot A_1\left(1\right)(X)\\&=X\cdot X+1\cdot X=X^2+X.    
    \end{aligned}
    \]
    Back to $A_4\left(0\right)(X)$ and $ A_8\left([0,0]\right)(X)$:
    \[\footnotesize
    \begin{aligned}
        A_4\left(0\right)(X)&=(1+X^2)^2 + (X^2+X)^2= 1+6X^2+X^4,\\
        A_8\left([0,0]\right)(X)&=(1\!+\!6X^2\!+\!X^4)^2\!=\!1\!+\!12X^2\!+\!38X^4\!+\!12X^6\!+\!X^8.
    \end{aligned}
    \]
    % Now, by enumerating the rest of the cosets, we can find the entire weight distribution for the code:
    % \[\footnotesize
    % \begin{aligned}
    %     A_8\left(0,0,0,0,0,0\right)(X)&=1+2 X^4+X^8\\
    %     A_8\left(0,0,0,1,0,1\right)(X)&=4X^4\\
    %     A_8\left(0,0,0,1,0,0\right)(X)&=4X^4\\
    % \end{aligned}
    % \]    
    Thus, $A_N(\A)(X)=1+12X^2+38X^4+12X^6+X^8.$ 
    
\end{example}

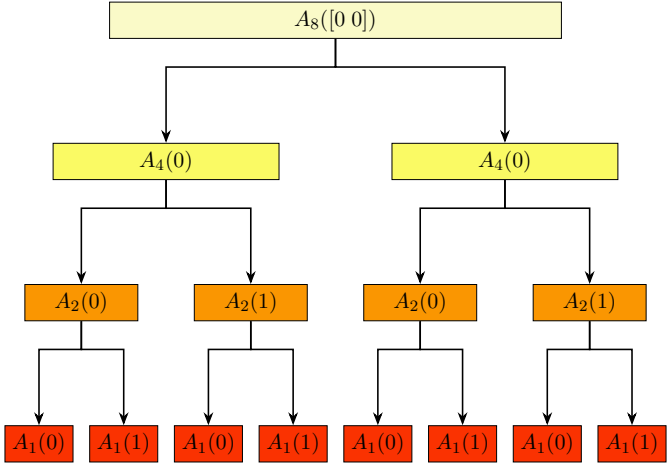
\begin{figure}[h]
    \centering
    \resizebox{1\columnwidth}{!}{%
        % Creating the TikZ picture for coset decomposition
        \begin{tikzpicture}[
        % Defining styles for nodes and edges
        level/.style={rectangle, draw, minimum height=0.5cm, align=center},
        arrow/.style={-Stealth, thick},
        level distance=2.5cm,
        sibling distance=3cm
        ]
        
        % Level 1: C_8 (width proportional to 8)
        \node[fill=c2,level, minimum width=8cm] (C8) at (0,0) {$A_8([0\;0])$};
        
        % Level 2: C_4 after u_1 (i=1 is odd, apply eq:enum_odd)
        \node[fill=c4,level, minimum width=4cm] (C4_0) at (-3,-2.5) {$A_4(0)$};
        \node[fill=c4,level, minimum width=4cm] (C4_1) at (3,-2.5) {$A_4(0)$};
        
        % Level 3: C_2 after u'_0 for N=4 (i=0 is even, apply eq:enum_even)
        \node[fill=c8,level, minimum width=2cm] (C2_00) at (-4.5,-5) {$A_2(0)$};
        \node[fill=c8,level, minimum width=2cm] (C2_01) at (-1.5,-5) {$A_2(1)$};
        \node[fill=c8,level, minimum width=2cm] (C2_10) at (1.5,-5) {$A_2(0)$};
        \node[fill=c8,level, minimum width=2cm] (C2_11) at (4.5,-5) {$A_2(1)$};
        
        % Level 4: C_1 after u''_0 for N=2 (i=0 is even, apply eq:enum_even)
        \node[fill=c10,level, minimum width=1cm] (C1_000) at (-5.25,-7.5) {$A_1(0)$};
        \node[fill=c10,level, minimum width=1cm] (C1_001) at (-3.75,-7.5) {$A_1(1)$};
        \node[fill=c10,level, minimum width=1cm] (C1_010) at (-2.25,-7.5) {$A_1(0)$};
        \node[fill=c10,level, minimum width=1cm] (C1_011) at (-0.75,-7.5) {$A_1(1)$};
        \node[fill=c10,level, minimum width=1cm] (C1_100) at (0.75,-7.5) {$A_1(0)$};
        \node[fill=c10,level, minimum width=1cm] (C1_101) at (2.25,-7.5) {$A_1(1)$};
        \node[fill=c10,level, minimum width=1cm] (C1_110) at (3.75,-7.5) {$A_1(0)$};
        \node[fill=c10,level, minimum width=1cm] (C1_111) at (5.25,-7.5) {$A_1(1)$};
        
        % Drawing arrows with exact indices
        % i=1 (odd, eq:enum_odd)
        \draw [arrow] (C8.south) -- ++(0,-0.5) -| (C4_0.north); %node[pos=0.25, left, font=\footnotesize] {$u_1=0$};
        \draw [arrow] (C8.south) -- ++(0,-0.5) -| (C4_1.north); %node[pos=0.25, right, font=\footnotesize] {$u_1=1$};
        
        % Point where the two lines go south
        %\coordinate (branch_point) at ($(C8.south) - (0,1)$);

        %\node at ($(C8.south) - (0,1.25)$) [circle, fill=white, draw=black, inner sep=1pt] {\footnotesize$\times$};

        % i=0 for N=4 (even, eq:enum_even)
        \draw[arrow] (C4_0.south) -- ++(0,-0.5) -| (C2_00.north); %node[pos=0.25, left, font=\footnotesize] {$u'_0=0$};
        \draw[arrow] (C4_0.south) -- ++(0,-0.5) -| (C2_01.north); %node[pos=0.25, right, font=\footnotesize] {$u'_0=1$};
        \draw[arrow] (C4_1.south) -- ++(0,-0.5) -| (C2_10.north); %node[pos=0.25, left, font=\footnotesize] {$u'_0=0$};
        \draw[arrow] (C4_1.south) -- ++(0,-0.5) -| (C2_11.north); %node[pos=0.25, right, font=\footnotesize] {$u'_0=1$};
        
        % i=0 for N=2 (even, eq:enum_even)
        \draw[arrow] (C2_00.south) -- ++(0,-0.5) -| (C1_000.north); %node[pos=0.25, left, font=\footnotesize] {$u''_0=0$};
        \draw[arrow] (C2_00.south) -- ++(0,-0.5) -| (C1_001.north); %node[pos=0.25, right, font=\footnotesize] {$u''_0=1$};
        \draw[arrow] (C2_01.south) -- ++(0,-0.5) -| (C1_010.north); %node[pos=0.25, left, font=\footnotesize] {$u''_0=0$};
        \draw[arrow] (C2_01.south) -- ++(0,-0.5) -| (C1_011.north); %node[pos=0.25, right, font=\footnotesize] {$u''_0=1$};
        \draw[arrow] (C2_10.south) -- ++(0,-0.5) -| (C1_100.north); %node[pos=0.25, left, font=\footnotesize] {$u''_0=0$};
        \draw[arrow] (C2_10.south) -- ++(0,-0.5) -| (C1_101.north); %node[pos=0.25, right, font=\footnotesize] {$u''_0=1$};
        \draw[arrow] (C2_11.south) -- ++(0,-0.5) -| (C1_110.north); %node[pos=0.25, left, font=\footnotesize] {$u''_0=0$};
        \draw[arrow] (C2_11.south) -- ++(0,-0.5) -| (C1_111.north); %node[pos=0.25, right, font=\footnotesize] {$u''_0=1$};

        \end{tikzpicture}
    }
    \caption{Enumeration via coset decomposition in Example \ref{ex:recursive_enum}.}
    \label{fig:coset_decomposition}\vspace{-10pt}
\end{figure}
%%%%%%%%%%%%%%%%%%%%%%%%%%%%%%%%%%%%%%%%%%%%%%%%%%%%%%%%%%%%%%%%%%%%%%%%
\section{Symmetry in Weight Distribution}\label{sec:code_symmetry}
%{\color{red}Reviewer 1 suggested to remove this part since it is much simpler than the rest of the sections.}
A binary linear code $\C(\A)$ of length $N$ is said to have a symmetric weight distribution if the number of codewords of weight $w$, $A_w$, satisfies the condition:
$$
A_w=A_{N-w} \quad \text { for all } 0 \leq w \leq N .
$$
%This symmetry implies that the weight distribution of the code is invariant under the transformation $w \mapsto N-w$. %Such a property is often exhibited by self-dual codes or codes with specific structural properties, such as those invariant under certain permutations. 
%Here, we use a different approach to show this property. 
We know that the generator matrix of a polar code $(N, K)$ always includes row $\bg_{N-1}$, corresponding to the most reliable synthetic channel. Similarly, any decreasing monomial code contains the monomial $1$.  %We know that a binary linear code $\C$ of length $N$ has a symmetric weight enumerator if and only if the all-ones vector $\mathbf{1}$ is a codeword in $\C$. This is the case for any decreasing monomial codes (containing monomial $\mathbf{1}\in\I$), or equivalently any polar codes where all-one row in the last row in the generator matrix $\bG_N$ which is a codeword itself. 

%Proof: Since a polar code $\C$ is a linear code, for any codeword $c \in \C$, the vector $c^{\prime}=c+1$ must also be in $\C$. The codeword $c^{\prime}$ is the bitwise complement of $c$. 
If a codeword $\bc$ has weight $w$, %(meaning it has $w$ ones and $N-w$ zeros), 
then $\bc^{\prime}=\bc+\mathbf{1}$, where $\mathbf{1}$ is an all-one vector, will have weight $N-w$. %ones and $w$ zeros. 
Therefore,  $\w\left(\bc^{\prime}\right)\!=\!N\!-\!\w(\bc)$.
This creates a one-to-one mapping between codewords of weight $w$ and codewords of weight $N-w$; $A_{w}\!\rightarrow\!A_{N-w}$. It follows that the number of such codewords must be equal. Thus, the weight distribution is symmetric.
A linear code $\C$ with $\mathbf{1}\!\in\!\C$ is called \emph{self-complementary}.

\begin{example}\label{ex:slf-compl}
For the (8,6) polar code with $\A=[2,7]$, we analyze the weight distribution of its codewords. Initially, the $2^5=32$ codewords are formed from row combinations of the basis $\A\setminus\{7\}$. An additional 32 codewords are generated by considering these initial codewords added to an all-one codeword (corresponding to row $\mathbf{g}_7$). The resulting weight distributions are summarized in  Table~\ref{tab:weights}.
\begin{table}[htbp]
    \centering
    \small % Uses a smaller font size for the table
    \setlength{\tabcolsep}{3pt} % Reduces column separation (default is 6pt)
    \caption{Weight distribution of the Code (8,6) in Example~\ref{ex:slf-compl}}
    \label{tab:weights}
    \begin{tabular}{|c|ccccccccc|c|}
        \hline\rowcolor{c2}
        \textbf{Weight ($w$)} & 0 & 1 & 2 & 3 & 4 & 5 & 6 & 7 & 8 & \textbf{Total}\\
        \hline
        Without $\mathbf{g}_7$ & {1} & 0 & {\color{red}9} & 0 & {\color{brown}19} & 0 & {\color{green}3} & 0 & 0 & 32\\
        \hline
        With $\mathbf{g}_7$ & 0 & 0 & {\color{green}3} & 0 & {\color{brown}19} & 0 & {\color{red}9} & 0 & {1} & 32\\
        \hline
        \textbf{Overall} & 1 & 0 & 12 & 0 & 38 & 0 & 12 & 0 & 1 & 64\\
        \hline
    \end{tabular}\vspace{-10pt}
\end{table}
\end{example}

\textbf{Even-weight Codewords.} 
Each polar codeword $\bc$ is a modulo-2 sum of rows {of} $G_N$, $\bc = \sum_{i \in \H \subseteq \A} \bg_i$. Since the least reliable bit-channel (index 0) is always frozen (i.e., $0 \notin \A$), none of the contributing rows in $\H$ have odd weight. Consequently, the sum produces an even-weight codeword as the sum of any number of even-weight binary vectors is always an even-weight vector.
%\end{remark}

%%%%%%%%%%%%%%%%%%%%%%%%%%%%%%%%%%%%%%%%%%%%%%%%%%%%%%%%%%%%%%%%%%
\section{Outlook and Future Directions}
% While this tutorial covers the latest advances in characterization of weight structure of polar codes based on permutation group action which can be applied on Reed-Muller codes as well, and based on recursive decomposition of codewords, the following challenges for further advances remain. 
% The closed-form formulae in the literature are focused on the $\wm$-weight codewords \cite{bardet,rowshan2023formation}, weights in $[1.5\wm,2\wm)$ \cite{vlad1.5d,ye2024distribution,rowshan2024weight,dragoi2025weight}, and $2\wm$-weight codewords \cite{rowshan2024weight}, only for low/high-rate codes where we have the complete weight distribution using formulae. However, devising formulae for $2\wm$ for all-rate polar codes and for weights larger than $2\wm$ remains open. Furthermore, these formulae work on plain polar codes, which have elegant structure, and enumeration of pre-transformed polar codes such as CRC-polar codes and PAC codes using closed-form formula remain to be addressed.

{T}his tutorial gives an overview of the latest advances in characterizing the weight structure {and enumeration} of polar {and Reed–Muller} codes based on permutation group action. {Nevertheless, several challenges remain. }

Existing closed-form formulae address {codewords with minimum-weight $\wm$ \cite{bardet,rowshan2023formation}, weights in $[1.5\wm, 2\wm)$ \cite{vlad1.5d,ye2024distribution,rowshan2024weight,dragoi2025weight}, and weight $2\wm$} \cite{rowshan2024weight}, primarily for low- and high-rate codes where entire weight distribution can be derived. However, deriving closed-form expressions for $2\wm$-weight codewords {and beyond} across all code rates remains an open problem. Moreover, these existing formulae apply only to plain polar codes, which possess elegant algebraic structure. The {closed-form} enumeration of pre-transformed polar codes - such as CRC-polar codes and PAC codes, remains unaddressed. 
{Algorithmic enumeration of the entire weight spectrum} leveraging the recursive structure of polar codes \cite{Arikan,polyanskaya,yao,liu2024method} remains challenging. This is primarily due to the rapidly increasing time complexity as the code length {increases}. 
% Deterministic weight enumeration based on the recursive structure of polar codes \cite{arikan,polyanskaya,yao,liu2024method} remains a significant challenge, primarily due to the rapid growth in time complexity as the code length exceeds 256 bits. 
% Extending weight structure analysis to non-binary polar codes over $\mathbb{F}_q, q>2$ could enable applications in non-binary channels or quantum error correction. 
% Moreover, while the weight spectra's impact on the error correction performance is known using Union-Bhatachariyya bound, analytical investigation of their impact at low SNR regimes and in particular considering the role of synthetic channels' reliability is are promising avenues to advance the theoretical understanding of polar codes. 

% Lastly, given recent advances in the characterization of the weight structure of codewords and associated properties can be applied in various venues. The main venue is benefiting the code-design from the discovered properties. Key challenges include optimizing monomial (or equivalently channel index) selection to achieve desired weight distributions, such as maximizing minimum distance or minimizing low-weight codewords, while not sacrificing the overall reliability of the code \cite{rowshan2025towards} is challenging. This optimization has to come at low-complexity to make it suitable for standardization. Additionally, enhancing polar codes for code-based cryptography by increasing minimum distance or integrating with other primitives could bolster post-quantum security.

{Recent advances in understanding codeword weight structure open new opportunities for code design. A central challenge is optimizing monomial selection (i.e., the information set $\A$) to shape the weight distribution—maximizing minimum distance or reducing low‑weight codewords—without degrading reliability, as shown in \cite{rowshan2025towards}. Such optimization must also remain computationally efficient for practical usability, e.g. in code design workflows. %\vlad{ is a bit vague. It would help to specify whether “for standardization” means in practical implementations, in standards bodies, or in code design workflows.}
% Recent progress in characterizing the weight structure of codewords presents numerous application opportunities. Foremost is the potential to inform code design based on these discovered properties. A key challenge lies in optimizing monomial selection (equivalently, information set $\A$) to engineer desired weight distributions, such as maximizing the minimum distance or minimizing low-weight codewords, without significantly compromising overall code reliability, as demonstrated in \cite{rowshan2025towards}. This optimization must be computationally efficient to align with standardization requirements. %The work in \cite{rowshan2025towards} utilizes the weight contribution of maximum-degree monomials and reliability partial order in the form of a diagram to enhance the rate-profile of a given 5G polar code, balancing between the weight distribution and reliability in a simple way. 

% Furthermore, %incorporating these techniques to enhance polar codes for code-based cryptography, by boosting minimum distance or integrating with complementary cryptographic primitives, may significantly improve post-quantum security. Moreover, 
% extending weight structure analysis to non-binary polar codes over $\mathbb{F}_q$, where $q > 2$, opens promising avenues for deployment in non-binary communication channels and quantum error correction. 
Finally, while the impact of weight spectra on asymptotic error correction performance is well-established through the {Union-Bhattacharyya bound}, deeper analytical investigation by considering the { weight contribution of individual synthetic channels} as an influencing factor, could {enhance} the understanding of polar codes.

%%%%%%%%%%%%%%%%%%%%%%%%%%%%%%%%%%%%%%%%%%%%
%\section*{Reproducible Research Resources}
\section*{Software and Data}
{The GitHub repository at \cite{Rowshan2025weightGITHUB}  includes the scripts implementing the enumeration methods, instructions, and sample inputs for reproducing the results in this article.}

\printbibliography

\begin{IEEEbiography}[{\includegraphics[width=1in,height=1in,clip,keepaspectratio]{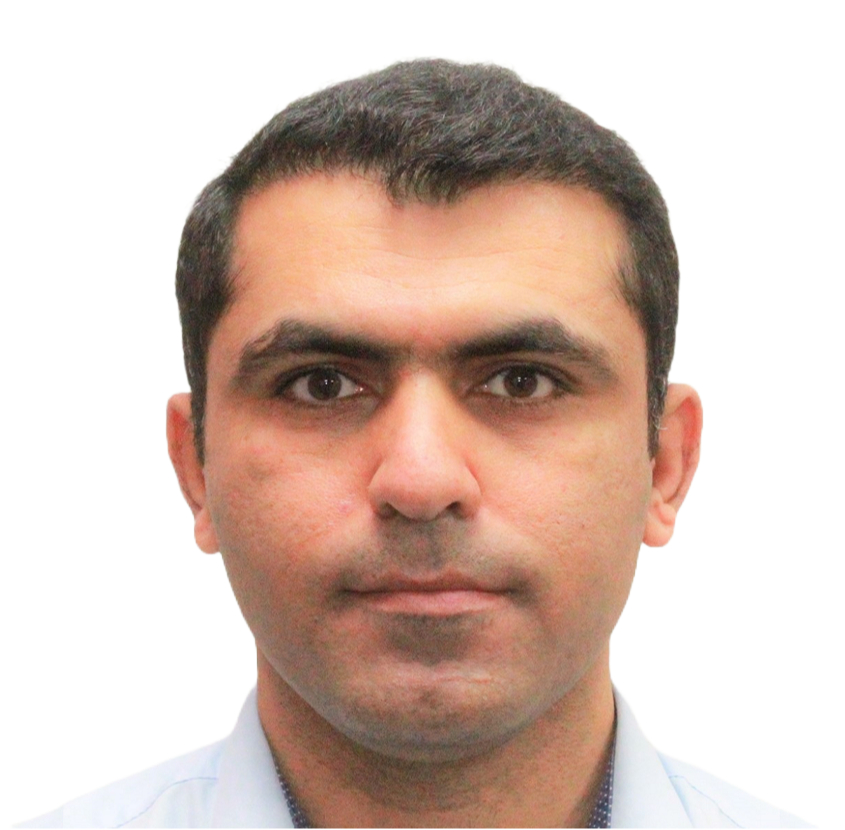}}]%
{Mohammad Rowshan } (S'13-M'22) is an Engineering ECA Fellow %in the School of Electrical Engineering and Telecommunications 
at the University of New South Wales (UNSW), Sydney, Australia. He received his B.Eng. (Hons) in Electrical Engineering from the University of Nottingham in 2015 (ranked 1), his M.Sc. in Integrated Circuit Design Engineering from the Hong Kong University of Science and Technology (HKUST) in 2016, %where he was awarded the Excellent Student Scholarship and the Arthur and Louise May Scholarship for Young Engineers; 
and his Ph.D. in Electrical Engineering from Monash University in 2021. %His research interests include classical and quantum error correction,  communication theory, and hardware architecture design.
\end{IEEEbiography}
\vskip -2\baselineskip plus -1fil 
\begin{IEEEbiography}[{\includegraphics[width=1in,height=1in,clip,keepaspectratio]{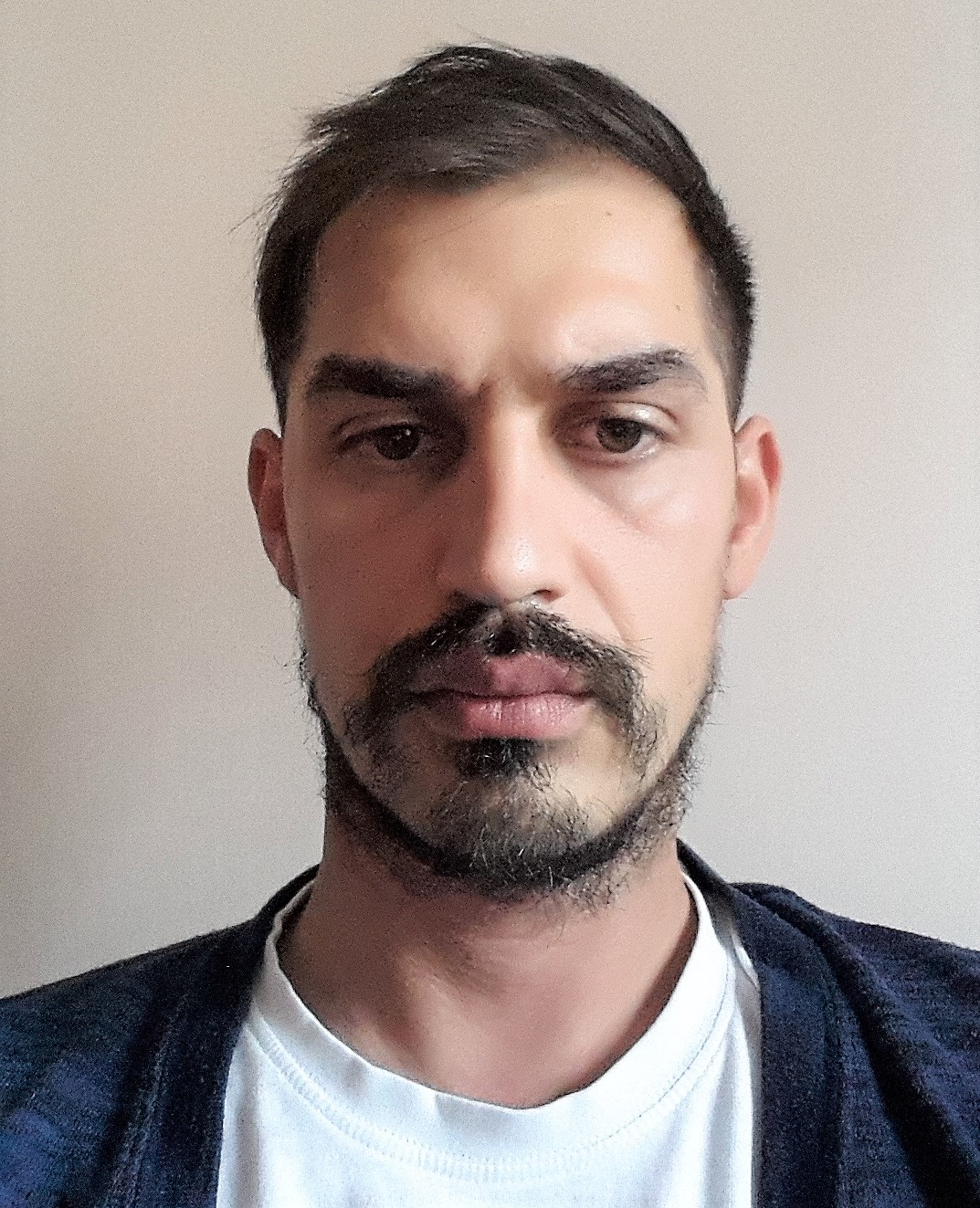}}]{Vlad Dragoi }
    received the B.Sc. degree in mathematics and the M.Sc. degree in applied mathematics, both from the UCBL, France, in 2011 and respectively 2013, and the Ph.D. degree in computer science from the University of Rouen Normandy, Rouen, France, in 2017. 
Since 2018, he joined the “Aurel Vlaicu” University of Arad (UAV), Arad, Romania, where he is currently Associate Professor. He was recipient of a national research grant for young researchers and a member of several research grants, both EU and national. %His research interests include post-quantum cryptography (with a focus on code-based cryptography), discrete mathematics applied to error correcting codes, and network reliability. 
%Dr. Dragoi received a doctoral scholarship from the University of Rouen-Normandy from 2013 to 2016, and is the recipient of two IEEE best paper awards.
\end{IEEEbiography}

\end{document}